# Active Protothrusts and Fluid Highways: Seismic Noise Reveals Hidden Subduction Dynamics in Cascadia

**Authors:** Maleen Kidiwela[1]*, Marine A. Denolle[2], William S. D. Wilcock[1], Kuan-Fu Feng[2,3]

**Affiliations:**

[1]School of Oceanography, University of Washington, Seattle, WA, 98105, USA.

[2]Department of Earth and Space Sciences, University of Washington, Seattle, WA, 98105, USA.

[3]University of Utah, Salt Lake City, UT, 84112, USA.

*Corresponding author. Email: seismic@uw.edu

**Abstract:**
**Complex interactions between strain accumulation, fault slip, and fluid migration influence shallow subduction zone dynamics. Using a decade of continuous ambient seismic data from Cascadia's seafloor observatories, we identified distinct regional variations in subduction dynamics. Northern Cascadia exhibits a fully locked megathrust with persistent strain accumulation, while central Cascadia displays a slow slip event on protothrusts and rapid fluid migration along fault systems in the overriding plate. Effective fluid transport through the décollement and the Alvin Canyon Fault likely modulates the earthquake behavior but does not cause slow slip events on the megathrust and likely stabilizes large earthquakes, promoting rupture arrest.**

**Teaser**

Long-term offshore seismic monitoring reveals strain and fluid migration patterns linked to Cascadia earthquake potential.

MAIN TEXT

## Introduction

The mechanical behavior of megathrusts is crucial for understanding and anticipating tsunami generation and earthquake hazards. The shallow subduction faulting structure affects both short-term earthquake rupture (*1*) and long-term accretionary wedge development (*2*). Active geophysical surveys (*3*) provide precise structural data, but most geophysical surveys are onshore, preventing direct observation of seismic activity (*4*). Active faults may behave seismically (with earthquakes), quasi-seismically (with slow earthquakes and tremors), or aseismically through creep. The absence of seismicity and lack of frictional data led to the belief that shallow seismic creep is ubiquitous (*5*). However, trench-rupturing earthquakes and new geodetic surveys reveal a slip rate deficit, likely due to frictional locking or stress shadowing (*6*). Slow slip and tremor downdip of the megathrust are observed worldwide (*7*), with mechanisms proposed to enhance this, such as fluid migration from mineral phase transitions (*8*). Shallow tremors and low-frequency earthquakes sometimes accompany slow slip events, as seen in the Nankai subduction zone, Japan (*9*), and the Guerrero segment of the Mexican subduction zone (*10*). Determining whether the shallow megathrust and upper-plate faults are locked, partially slipping, or creeping is challenging without near-trench seismic and geodetic observations (*11*).

The active Cascadia Subduction Zone (CZS) exhibits unique characteristics: low background seismicity and downdip episodic tremor and slip (ETS) (*12*). However, its history of large and tsunamigenic earthquakes (*13*) suggests complex deformation mechanisms in the shallow megathrust that remain unexplained. Variations in the downdip ETS and modeled megathrust slip-rate deficit (*14*) may indicate that northern Cascadia is fully locked to the deformation front (Fig. 1), while central Cascadia may be partially locked. The central Cascadia megathrust splay faults are seaward-verging and intersected by west-northwest-striking, left-lateral strike-slip faults, such as the Alvin Canyon Fault (ACF, Fig. 1) (*15*). Recent seismic imaging studies have confirmed the ACF as a major structural discontinuity and crustal tear across the margin, which shifts the slab depth deeper by ~1.5 km along strike (Fig. 1) (*3*). Additionally, central Cascadia has thicker underthrusting sediments and a shallower décollement compared to northern Cascadia (*16*), likely reducing fault coupling. Fluids play a key role in both ETS and shallow earthquakes (*17*), with deep-origin fluids released to the seafloor (*18*, *19*). These fluids are generated at the mantle wedge and released updip during ETS (*20*). Despite extensive research, the migration pathways, whether through hydraulically connected faults or more complex routes in the upper plate, remain uncertain (*16*, *20*, *21*). This uncertainty holds significant implications, as it may directly influence the behavior of the next major earthquake in Cascadia.

The lack of continuous geodetic monitoring in the shallow subduction zone has limited our ability to assess the offshore seismic activity. Recent advances in seismic monitoring, particularly the use of temporal changes in seismic velocity (dv/v), provide a powerful proxy for measuring subsurface strains (*22*). The approximation that volumetric strain $\varepsilon = -(1/\beta)*dv/v$, with $\beta$ typically around ~ $10^3$-$10^4$ across various regions, units, and materials (*23–26*), seismic frequencies, and depth, allows us to estimate strain values from dv/v. Continuous seismometer recording of ambient seismic waves now enables frequent, high-resolution strain measurements (*27*), once thought unattainable. Using this breakthrough method and a decade of data from broadband seafloor seismometers connected by the Ocean Observatory Initiative Regional Cabled Array in central Cascadia and the Ocean Networks Canada NEPTUNE cabled observatory in northern Cascadia, we gain new insights into offshore strain dynamics, opening opportunities to study fault activity in real-time using dv/v.

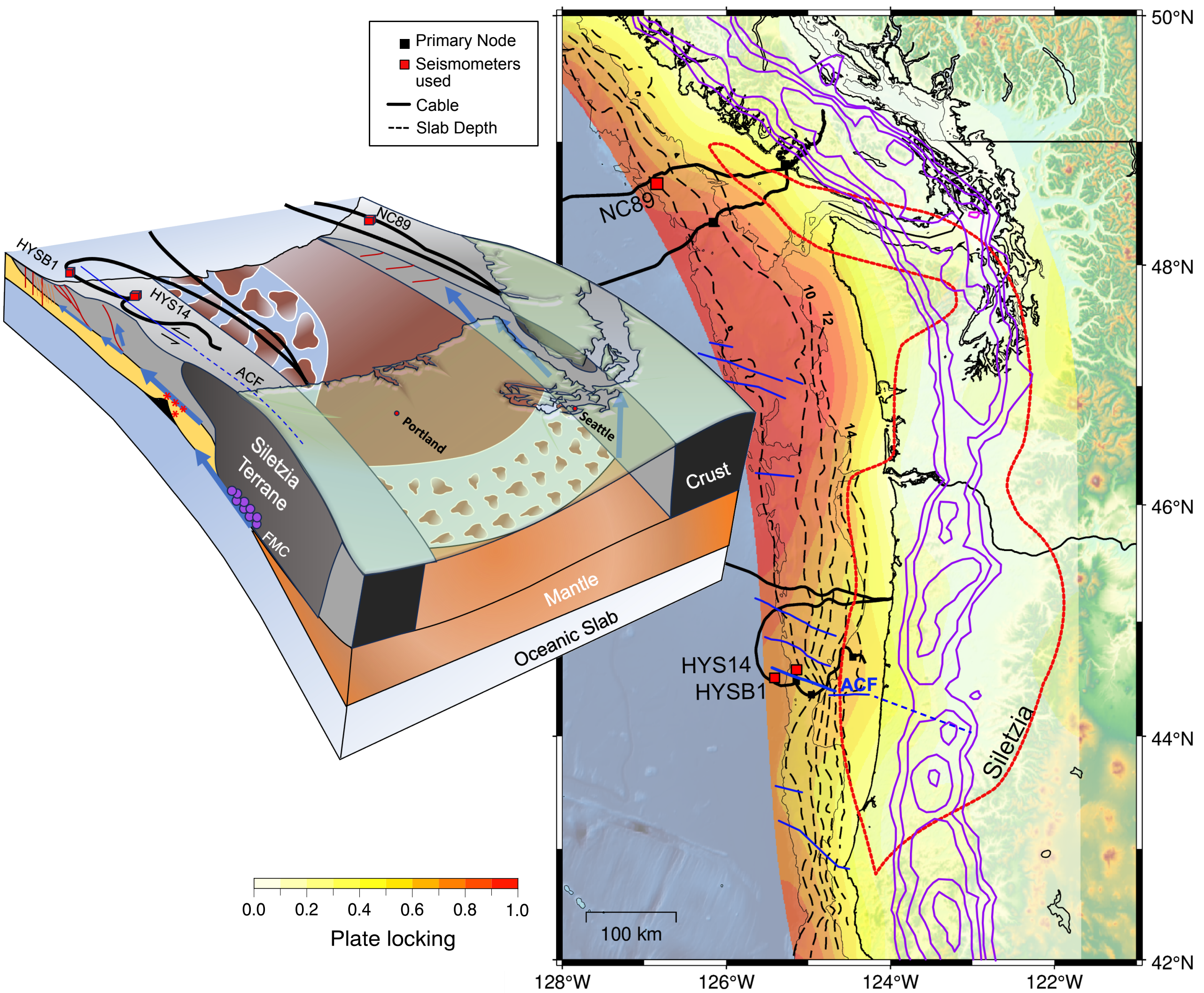

**Figure 1: Map and schematic of the subduction-zone system in Central and Northern Cascadia.** The map shows a locking model for Cascadia (*28*), with red contours marking the Siletzia terrane and purple contours representing tremor density (logarithmically spaced 30 to 2,430 per 100 km). Dashed contours indicate plate interface depths (*3*), and blue lines show strike-slip faults (*15*). The dashed blue line extends the Alvin Canyon Fault (ACF), where tremor density decreases. Geological units are labeled: the Siletzia Terrane (dark gray), accretionary wedge (light gray), upper mantle (orange), oceanic slab (light blue), forearc crust (light gray), and underthrust sediments in central Cascadia (yellow). The cross-section shows fluid migration with blue arrows, splay faults and protothrusts in red, and the ACF in blue. The observatory cables and studied stations are shown as black lines and red boxes. Stress barriers are marked in brown, and seismicity related to the black seamount is shown by red stars. Downdip tremor is indicated by purple markers, matching small, locked patches.

## Results

We analyzed single-station cross-component ambient noise interferometry at the northern Cascadia Clayoquot Canyon station (NC89) and at the central Cascadia stations Hydrate Ridge (HYS14) and Slope Base (HYSB1) (Figure 1). In central Cascadia, we also conducted interstation ambient noise interferometry between these stations.

**Northern Cascadia** shows a persistent, quasi-linear increase in seismic velocity, with a dv/v trend of +0.038% per year over 13 years in the 1-3 Hz band (Fig. 2). Assuming $\beta = 10^3$-$10^4$, we calculated a volumetric strain rate of approximately 0.04-0.4 µstrains/year. This is consistent with borehole pressure data from Hole 1364A in proximity to NC89 at Clayoquot Canyon, which shows compressional volumetric strain rate of 0.12 µstrains/year (*29*), and with the linear geodetic strain rate of 0.4 µstrains/year from a 4.1 cm/year convergence rate over 100 km; although linear convergence represents only part of the total volumetric strain. This aligns well with the full locking inferred from onshore geodetic data in northern Cascadia, where megathrust locking causes compaction of the shallow accretionary wedge sediments, increasing seismic velocities in the subsurface.

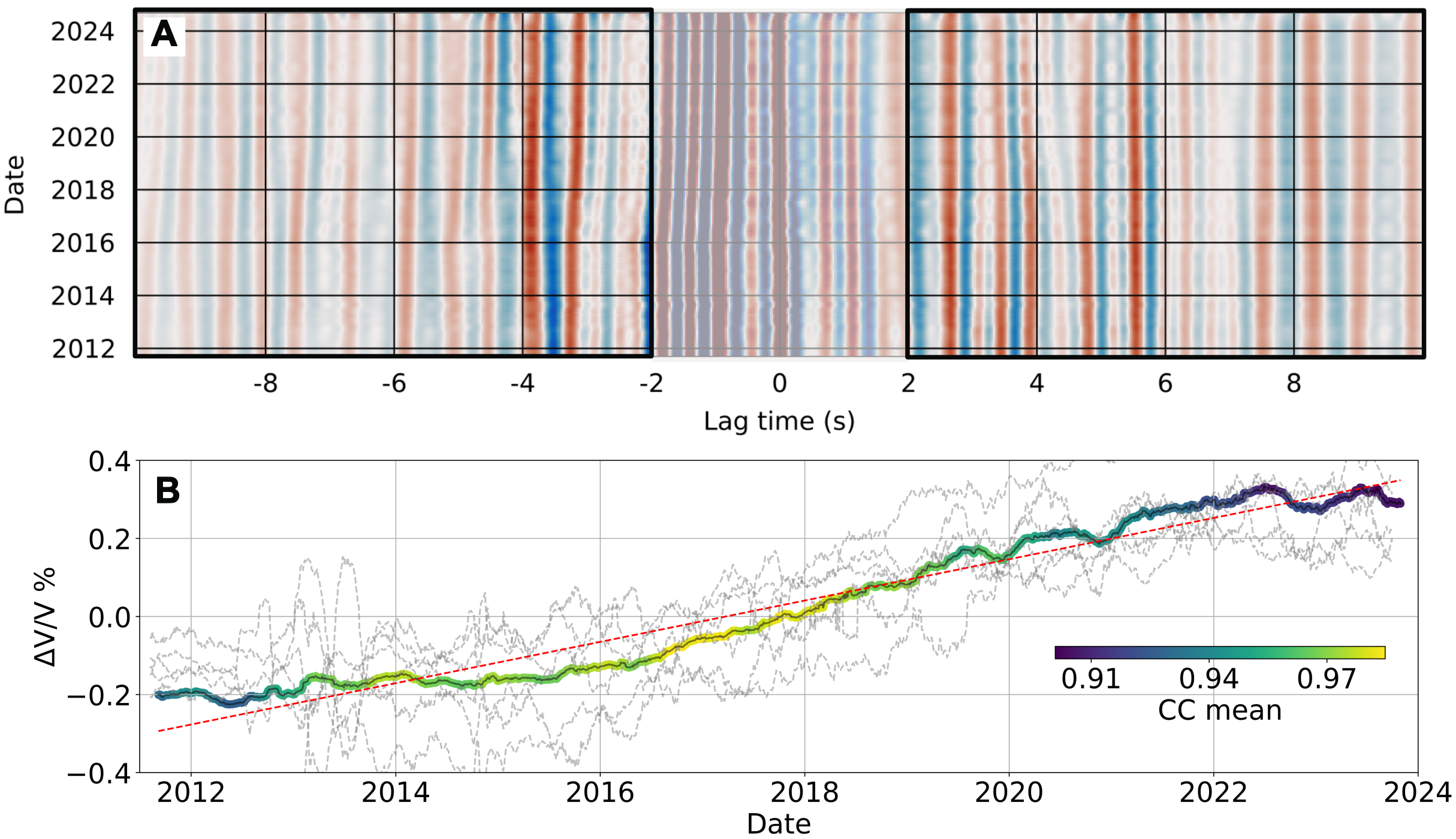

**Figure 2: Long-term dv/v in northern Cascadia.** (**A**) Single-station cross-correlation between HHZ and HHE components at NC89, bandpass filtered from 1-3 Hz and smoothed with a 60-day running mean, with the coda window for dv/v measurements highlighted. (**B**) dv/v measurements, showing individual causal and anticausal time series for ZE, NE, ZN cross-components in gray, with the weighted average represented by a color-coded line and the average correlation coefficient. The absence of sudden changes in (A) and the steady increase in dv/v in (B) indicate continuous strain accumulation.

The velocity changes in **central Cascadia** differ significantly from those in northern Cascadia, with two key events. First, a drop in seismic velocity occurred on 17 July 2016, lasting about two months (Fig. 3). Temporal changes in dv/v are typically observed during earthquakes (*22*). Still, in this case, the changes are not instantaneous (Fig. S1), and do not follow standard earthquake relaxation models (*30*). Since this event was not observed at Hydrate Ridge, we rule out the possibility of a slow earthquake on the megathrust or the ACF. Submarine landslides are also unlikely, as they typically increase seismic velocity (positive dv/v) due to the added load of sediments sliding down to the slope base. Instead, reflection surveys (*16*) show 45-degree dipping protothrusts beneath the station, with approximately $a$ = 3 km fault widths. A dv/v drop of 0.2% yields a maximum volumetric strain drop of $\varepsilon$ = 2 μstrains. The maximum displacement of a half-crack on a traction-free boundary $Dmax = 4(1-\nu)a\varepsilon \sim 1$ cm *(50),* where the Poisson ratio $\nu = 0.5$ is consistent with elevated pore pressure in shallow sediments (*16*). The resulting displacement is consistent with those of known slow slip events (*31*).

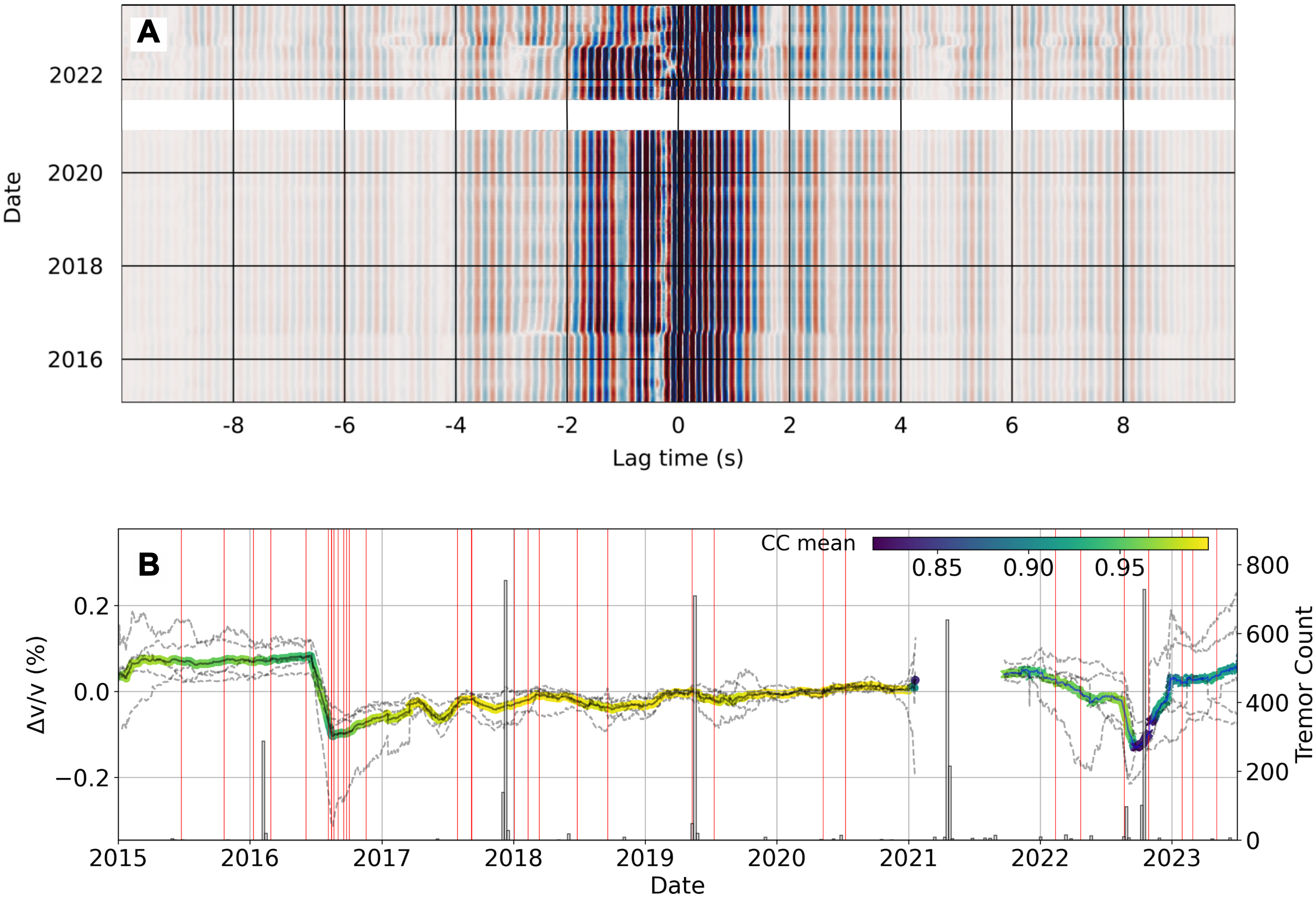

**Figure 3: dv/v suggests a slow slip and episodic fluid expulsion in central Cascadia protothrusts. (A)** Cross-correlation function between BHZ and BHE components at HYSB1. **(B)** dv/v at HYSB1 in the 3-5 Hz range, averaged over causal and anti-causal windows of ZE, ZN components and causal window of NE cross-components in dashed grey lines. Red lines mark shallow tectonic tremor timing (*32*), and black histograms show deep tectonic tremor timing at 44.4˚N-44.6˚N (PNSN, binned at 7 days, (*33*). Unlike northern Cascadia, where strain builds monotonically, the Pacific Plate shows an asymmetric 2-month dv/v drop with partial recovery, interpreted as slow slip on the protothrust. Several transient pulses in 2019 and 2022 suggest slip-free strain changes, likely due to fluid migration.

A second event type was observed in 2022 on the Slope Base, marked by a gradual decrease in seismic velocity followed by a slow recovery over several months (Fig. 3). Although only the HYSB1 station was operational in 2022, similar events were detected in 2019 at both HYS14 and HYSB1 by focusing on the period from 2017 to 2021. Unlike earthquake-induced changes, this gradual drop and recovery does not indicate strain released through fast fault slip. During these transients, the wavefield correlation coefficient (CC) decreases (Fig. S7), suggesting a temporary change in scattering properties (*21*). We propose that these events are pulses of elevated pore pressure, likely driven by fluids. Similar dv/v and CC transients have also been documented during fluid-injection operations (*34*). The most notable results are the lag times between the peaks of these pulses at different sites and frequencies. In the 1-3 Hz, a 34-day lag is observed between HYS14 (megathrust side) and HYSB1 (protothrust side), implying a horizontal propagation velocity of 0.58 km/day. The inter-station correlations show that the phenomenon occurs both at and between the stations (Fig. S5). At higher frequencies (3 - 5Hz), which are more sensitive to shallower depths, the lag times are longer–69 days at Slope Base and 57 days at Hydrate Ridge. These differences imply upward fluid migration velocities of 4 m/day at Slope Base (HYSB1) and 9 m/day at Hydrate Ridge (HYS14). Another key feature is that the pulse narrows below HYS14 (lasting 4 months) at greater depths but broadens as it moves past the deformation front at Slope Base (6 months) and into shallow sediments. The broadening suggests diffusion. The primary fluid migration pathways between Hydrate Ridge and Slope Base are the décollement (*16*) and ACF (*18*), while splay faults, ACF and the proto-thrusts (*16*) play a crucial role in upward fluid transport. Further evidence for fluid migration is shown by the drop in the lagged correlation coefficient and temporal variations, indicating scattering changes likely induced by fluids (Fig. S9).

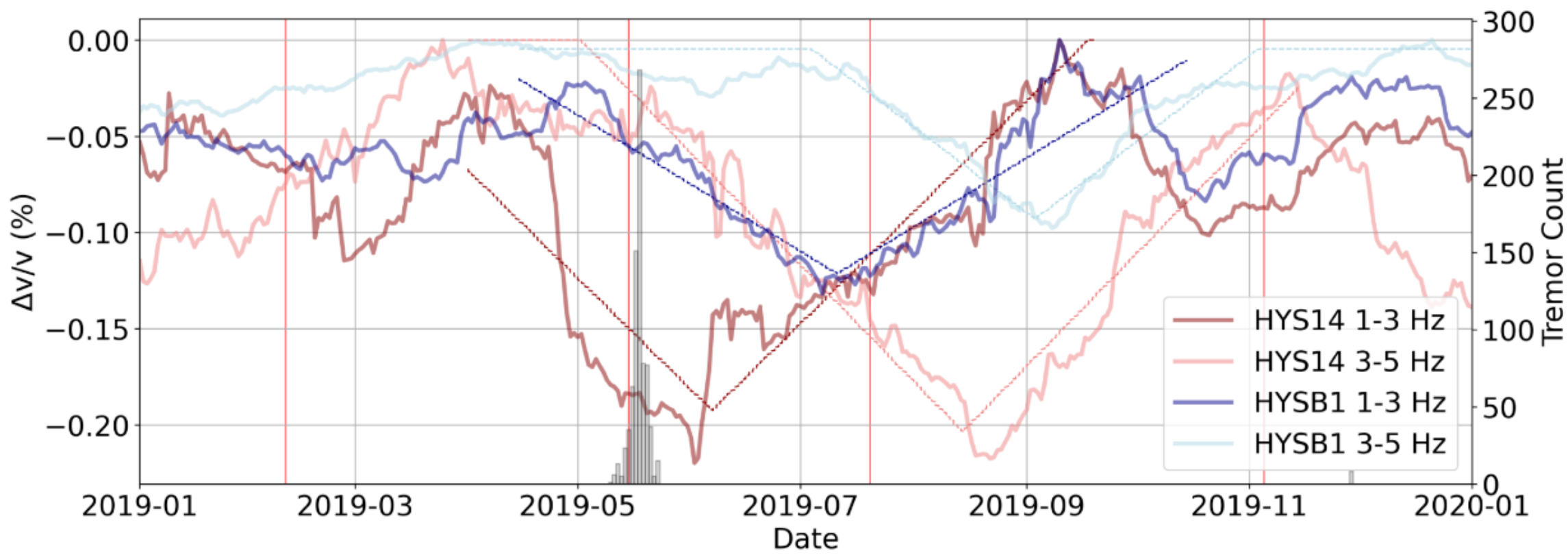

**Figure 4: Multi-component weighted dv/v observations associated with fluid pulse migration during the 2019 ETS event.** dv/v for Hydrate Ridge (HYS14) and Slope Base (HYSB1) stations in 2019 at 1-3 Hz and 3-5 Hz frequency bands, with deeper and shallower sensitivity *(50)*. Red represents the downdip site at Hydrate Ridge (HYS14), one on the upper plate, and blue represents the updip site at Slope Base (HYS1B), past the deformation front on the downgoing plate. Dashed lines are triangle functions fitted to velocity drops. Lag times are intervals between the lowest points of these functions.

Finally, another key finding is the timing of downdip tremor with the onset of the 2019 and 2022 events (Figs. 3 and 4). Hydromechanical models predict that solitary pore pressure waves, without fluid transport, could propagate updip at approximately 90 km/hr (*35*) along the décollement and are consistent with speeds of rapid tremor streaking in Cascadia (*36*). This may explain the synchronization between deep ETS and fluid pulses at Hydrate Ridge (Fig. 4) in our data. While downdip ETS occurs regularly, the shallow response is more erratic–no dv/v drop was detected during the 2016, 2018, and 2021 ETS events. Hierarchical clustering of tremor spatial density *(50)* shows that only the highest radiative energy released, spatially connected tremor episodes with large tremor counts coincided with dv/v drops (Figs. S8-S10). These results provide the first in situ measurements of shallow fluid release during ETS events in Cascadia, offering new insights into the interaction between fluid migration and seismogenic processes.

## Discussion

The long-term trends dv/v between northern and central Cascadia align with variations in geodetic models of slip-rate deficit and underlying mechanisms at play. In northern Cascadia, the consistent increase in seismic velocity and its magnitude suggests a high locking ratio at the megathrust, consistent with observations of a fully locked margin up to the deformation front. This interpretation is supported by relatively low offshore seismicity (*37*), a steady increase in borehole pore pressure (*29*), and higher offshore locking ratios derived from onshore geodetic data (*28*). In contrast, central Cascadia shows both lower strain accumulation and the presence of transient phenomena, suggesting a lower locking ratio. These differences are also apparent in the lowest frequency band (0.1-0.3 Hz), with dv/v rates of 0.006%/year in northern Cascadia and 0.003% in central Cascadia (Fig. S4). When interpreted in terms of locking rate, the locking ratio in central Cascadia is approximately half that of northern Cascadia. This observation is consistent with models of slip-rate deficit inferred from lower coastal uplift rates (*38*) and onshore geodetic data (*14*, *28*). The difference in locking ratios may also be influenced by factors such as the subduction of a seamount (*3*), the underthrusting of fluid-rich sediments (*16*), higher seismicity rates compared to other parts of Cascadia (*37*), and the presence of very low-frequency earthquakes (*39*).

ETS events in subduction zones are suspected to trigger the upward migration of deep fluids (*40*). At 30–40 km depth, near the forearc mantle corner, the dehydration of subducted metabasalt releases fluids that accumulate and build up pore pressure beneath an impermeable seal, as indicated by low-velocity, high Poisson's ratio anomalies in the oceanic crust (*17*, *41–44*). A "fault-valve" mechanism, which couples fault slip, the evolution of near-fault permeability, and fluid flow driven by pressure gradient (*20*, *45*), may explain the temporal variations in seismic properties observed in the ETS zone (*46*). We propose that synchronizing deep ETS with fluid-related transients in dv/v points to the episodic release of deep fluids through the seafloor, as also hypothesized in other subduction zones like Nankai (*21*, *40*). Our hierarchical clustering of the deep ETS events in Cascadia between latitudes 44.4˚N and 44.7˚N reveals that ETS clusters occurring concurrently with dv/v drops in 2019 (Fig. 4) and 2022 (Fig. 3) are tightly grouped (Figs. S8-S10). The tremor density analysis shows that these clusters are denser and interconnected, with many ETS events (Fig. S10). This suggests that specific characteristics of ETS may be modulating a pore pressure wave that propagates from the ETS zone to the seafloor and along the décollement.

The permeability structure of the upper plate plays a crucial role in the propagation of solitary pore-pressure waves or in the migration of fluids themselves during ETS and co-seismic ruptures (*47*). When pore pressure reaches a critical level, trapped fluids are often released during slow slip events due to dilatancy (*35*, *48*), typically coinciding with low-frequency earthquakes. These fluids begin to migrate upward, either through the overriding plate if it is permeable or along a megathrust channel (*40*, *46*, *49*). The more permeable upper plate in northern Cascadia allows fluids to migrate through the upper plate itself, as observed in Hikurangi, New Zealand. In contrast, in central Cascadia, where the upper plate is less permeable due to the presence of Siletzia terrane (*50*), we propose that fluids migrate along the décollement, similar to those in the Nankai Trough (*40*). The fluids that migrate may be stored in shallow regions up-dip edge of Siletzia (*50*). The solitary pore-pressure waves may or may not be related to these longer-term fluid migrations (43) as described in (29).

The Alvin Canyon Fault (ACF) is a crustal tear across the margin (*3*) that serves as another potential conduit for the observed fluid pulses during deep ETS events, as it is known to transport fluids from the décollement to the surface (*18*). Similar to the ACF, the Wecoma fault is another strike-slip fault north of the ACF that has been shown to be a high-angle fluid escape conduit, dewatering the megathrust décollement (*51*). Whether fluids move upward along the ACF or the décollement, both pathways indicate effective fluid transport, preventing the buildup of high pore pressure during coseismic events and thereby mitigating dynamic weakening effects such as thermal pressurization. The stabilization effect may explain why Hydrate Ridge has been a location where many partial margin ruptures have terminated (*13*). Additionally, the presence of ACF, Daisy Banks, and Wecoma strike-slip faults in this region likely locally inverts thrusting to seaward vergence (*51*) by facilitating the escape of highly overpressured fluids from the megathrust décollement.

**In Northern Cascadia**, magnetotelluric (*50*) and seismic imaging (*52*) studies reveal that the forearc crust is relatively conductive, exhibiting lower seismic velocities and a high Vp-Vs ratio, which indicates fluid-rich conditions. These characteristics suggest that fluids, potentially at lithostatic pressure, can migrate vertically within the overriding plate (*43*, *44*). Additionally, seismic velocity drops followed by rapid recovery during deep ETS events have indicated a temporal reduction in pore pressure and the breaching of a low-permeability material near the megathrust during slow slip events (*20*). The regular expulsion of fluids from the deep ETS zone into the upper plate likely prevents pore pressure buildup along the megathrust, thereby enhancing the fault clamping and locking (*53*). Furthermore, borehole pressure data from U1364A show no transient pressure change during slow slip events (*54*), which aligns with our observations of stable dv/v measurements in northern Cascadia.

The upper plate structure of **Central Cascadia** is dominated by the thicker, impermeable Siletzia Terrane, which contacts the plate interfaces and extends offshore into central Oregon. The ACF plays a key role in the shallow subduction system, acting as a barrier to the along-strike propagation of both fast and slow earthquakes (*3*). Fluids generated at depth become trapped up-dip and stored on the up-dip edge of Siletzia Terrane (*50*). The geological structure between the downdip ETS zone and the deformation front is complex, including a subducted seamount, land-vergence splay faults thought to be seismically active (*4*), and protothrusts interpreted as being fluid-saturated past the deformation front (*55*). Geochemical evidence shows that fluids from the frontal protothrust zone and Hydrate Ridge contain significant amounts of fluids originating from the décollement (*19*). We propose that hydraulic connectivity between the updip ETS zone and the frontal thrust, where pore pressure waves propagating from the ETS zone drive the flow of shallow stored fluids (*50*) up to the protothrusts along the decollement or ACF. Numerical models suggest that fault-valve instability, driven by cyclic permeability changes during slow slip events, triggers the propagation of pore-pressure waves (*35*). The décollement and the ACF are the primary fluid pathways, as indicated by negative polarity reflections at the base of the protothrusts (*16*, *55*, *56*). Although these pulses occur at a greater depth than the dv/v sensitivity shown in Fig. 4, our depth-integrated measurements, which have enhanced sensitivity at lower frequencies of dv/v, also contain transient dv/v drops, confirming that the source of these perturbations lies in the deeper crust (Fig. S6).

As the elevated pore pressure wave arrives in the shallow subduction zone, we propose that the high fluid pressure gradient induces flow along hydraulically connected pathways, such as décollement, splay faults, protothrusts, and cross-cutting strike-slip faults. The vertical migration is well explained by diffusion through narrow faults distributed across complex networks, including splay faults and protothrusts. We find a vertical diffusivity of ~$3.8\times10^{-4}$ $m^2$/s, consistent with hydraulic diffusivity values in Nankai and Costa-Rica *(50)*. Horizontal migration is similarly well explained by a flow in a 1 km-thick conduit with a hydraulic diffusivity of ~$6.3x10^{-2}$ (*53*), which aligns with expectations based on hydrogeological parameters (*57*) and the role of underthrusted sediments above the plate interface in fluid transport (*16*). Over time, the pore pressure in the seaward décollement may cause a breach, lowering the effective normal stress within the proto-thrusts and leading to slow slip.

Our transient dv/v signatures provide the first *in-situ* evidence of the dynamics and timescales associated with fluid transport in shallow subduction zones. Our observations are consistent with the possibility of a rapid pressure wave migrating from the ETS zone, which transitions into a slower fluid migration accommodated by flow in the primary conduits such as the décollement and the ACF and diffusion vertically to the seafloor through more distributed fault networks along protothrusts, splay faults, and cross-cutting ACF. We hypothesize that elevated pore pressure buildup at the base of protothrusts drains through intermittent slow slip and fluid diffusion. Over the past few decades, fluid transport has not been accompanied by slow slip on the shallow megathrust, nor has it altered the strain accumulation rate on the central Cascadia megathrust. The current locking ratio at 50% (*28*) (Fig. 1) implies that partial creep has proceeded uninterrupted by fluid transport phenomena during the instrumental record. We cannot infer that fluids are pressurized enough to trigger earthquakes on the splay faults or the ACF at this location. While this does not negate the possibility of their activation during large megathrust earthquakes (*4*), their role as fluid pathways likely reduces fluid-induced dynamic frictional weakening effects, stabilizes large ruptures, and acts as barriers to megathrust earthquakes in Cascadia. The upcoming expansion of the seafloor observatory will provide new opportunities to validate our interpretation further and enhance our ability to detect microseismicity that may otherwise be overlooked by the current network of two stations.

## Materials and Methods

### Ambient noise cross-correlation

Ambient Noise interferometric workflows are well-established, and we utilize open-source software in the programming language Julia (SeisNoise.jl (*58*) and SeisDVV.jl) as well as in Python.

First, station metadata is read from the appropriate StationXML files. These metadata include the instrument response, which is then removed from the raw data to convert the waveforms into physical units of ground motion. Next, the raw time series for each day is detrended, tapered, and initially bandpass-filtered within the frequency range of interest (0.01-10 Hz), ensuring the removal of long-term offsets and out-of-band noise. Amplitude clipping is applied to reduce the effect of high-amplitude transients, and whitening is applied to mitigate the influence of strong narrow-band signals. Finally, the cross-correlation is computed between pairs of channel data streams, with a 1800 s window length and a 100 s lag time, and the results are saved in files. Cross-correlations are stacked daily (or over another specified interval) to improve the signal-to-noise ratio and saved to a file. We perform the dv/v analysis over different frequency bands by applying bandpass filters (0.1-0.3, 0.3-0.6, 0.5-1.0, 1.0-3.0, 3.0-5.0 Hz). We then smooth the cross-correlations using moving average stacks (40-day or 60-day) with a 1-day step to enhance the stability of the correlation function. We further improve the cross-correlation using the denoising method, which involves combining the singular value decomposition (SVD) of a 2-D NCF with a Wiener filter (*59*). By applying the Wiener filter to the singular vectors, we maximized the coherency directly in the signal subspace before reconstructing the cross-correlation matrix. This global frequency-domain Wiener deconvolution is not a moving window smoother and does not artificially extend months-long signals. This method is essential for producing stable correlation functions in our ocean-bottom seismometer data, where incoherent signals resulting from non-uniform noise sources are readily present. We created dv/v time series, excluding values with correlation coefficients (CC) less than 0.8. After creating the dv/v time-series $n$ for each channel pair, $i$, we performed weighted averaging based on the correlation coefficient (Equation 1-2 obtained from (*23*)) to obtain the weighted dv/v time-series. Equation 3 calculates the mean correlation coefficient (*CCmean)*, over a specified number of channels ($N_{ch}$) shown in each dv/v time series (Figure 2 and 3)

$$CC_{sum} = \sum_{i=1}^{n} cc_i^2 \tag{1}$$

$$dv/v = \frac{1}{CC_{sum}} \sum_{i=1}^{n} cc_i^2 \, dv/v_i \tag{2}$$

$$CC_{mean} = \frac{1}{N_{ch}} \sum_{i=1}^{n} cc_i \tag{3}$$

Our investigation revealed that changing the stacking length, window length, and choice of reference cross-correlation significantly control the stability of dv/v observations. We performed a series of experiments to determine which parameters help produce a robust dv/v time series.

60-day stacks are ideal for observing long-term changes, but are not capable of observing transients that are shorter and smaller in magnitude. 40-day stacks are ideal for observing shorter transients. To increase the temporal resolution, the moving average stacking length needs to be changed from 40 days to 3 days. But this alone would not improve the stability of the time series. With the high noise power, the number of singular values must be manually reduced rather than using the Wiener filter. The 2016 asymmetric dv/v (drop + recovery) remains a consistent feature in 3-day stacks, albeit with increased scatter. Therefore, the slow slip is resolved at this frequency.

Separately, dv/v measurements are also affected by parameters such as the stretching window length, stacking length, choice of coda window, choice of reference, and width of the frequency bands. Use of different references other than limiting to a global reference is an important tuning parameter for seismic monitoring (*60*). The reference is affected by the strongest signals: the 2016 slow slip event is the largest of all signals and obscures the velocity changes of smaller transients that induce a phase shift. This is apparent in the muted visibility of the 2019 velocity drop in Figure 3. In order to emphasize the smaller events, we created a separate reference (2017-2021), excluding the 2016 slow slip event and reducing the correlation function moving average to 40 days (Fig. 4). We chose the stretching window based on the coda decay and the signal-to-noise ratio (SNR) of the correlation function. To achieve this, we computed the squared envelope of the correlation function to determine the coda decay window and its signal-to-noise ratio (SNR). We picked windows where the SNR is higher, and the coda decay rate is higher. This investigation showed that the dv/v time series are more stable when the window lengths are longer in lower frequencies and shorter in higher frequencies.

We experimented with frequency bands, including a 1 Hz bandwidth, in frequencies above 1 Hz. These bands yielded increased variance compared to a 2 Hz bandwidth. However, through an iterative approach, we settled on the hyperparameters listed in Table S1.

### Depth sensitivity analysis

We used the Ridge2Trench P-wave velocity models (*16*) for HYSB1 and HYS14. Based on the Vp/Vs values from the ridge-to-trench study, we calculated an estimated S-wave velocity profile for both locations. We used the open-source software: Disba (*61*) to solve for the depth sensitivity with depth for single station cross component correlation, which we ran for several frequency bands of interest. The script for running this analysis is available in a Python notebook (src/sensitivity_analysis.ipynb) within our main project repository.

We used a barycentric estimate of the depth sensitivity for each frequency, calculated using $H_d = \frac{\int Kz\,dz}{\int K\,dz}$, with $K$ the frequency-dependent depth sensitivity shown in Figure S2, to determine the depth of each frequency band. We decided to use this approach because most kernels consist of singular peak. For 0.1 Hz kernel with two peaks, we used the deeper more prominent peak rather than the shallower peak. Table S2 shows each depth estimate corresponding to the frequency. When calculating the distance of fluid migration, we calculated the difference between the lowest and highest depths of the two bands that were compared.

### Strain calculation

A volumetric strain rate of 0.12 μstrains/year (*29*) was calculated for the northern Cascadia subduction zone using the borehole 1364A. Borehole 1364A is located in proximity to the NC89 seismometer, which we used to assess seismic velocity variability. We used this strain value to optimize the crustal sensitivity β for the CSZ, based on our observed dv/v increase rate of 0.038% per year. This yielded a β value of 3.16 x $10^3$. In Nankai, a β value of 2.14 x $10^3$ could be calculated using a volumetric strain rate of 0.14 μstrains/year (*29*) and a dv/v increase of 0.03% per year (*21*). Therefore, these β values are consistent with each other even under the presence of large scale geological differences between Nankai and Cascadia.. We used the β value of 3.16 x $10^3$ for Cascadia to approximate strain values for the central Cascadia strain drop during the inferred slow slip event.

**Slip calculation**

In linear elasticity (shear mode):

$$\Delta\tau = G\,\varepsilon, \tag{3}$$

where, Δτ is the shear stress drop, G is the shear modulus (also denoted μ), ε is the net shear *strain* change ("strain drop") in the region of the fault/crack. In an infinite medium, the standard stress drop of a penny-shaped crack of radius a in an infinite isotropic medium is

$$\Delta\tau = G/\left[\,2\,(1-\nu)\right]\; D_{max} \tag{4}$$

where ν is Poisson's ratio and Dmax is the peak (center) slip. Therefore, Dmax = 2 (1−ν) a ε

If the same circular shear crack of radius a is cut in half by a free surface or boundary, its effective shear stress drop is about half that in the infinite-medium case (the crack is more compliant). Hence, $Dmax = \; 4\,(1-\nu)\,a\,\varepsilon$

**Full Waveform Modeling for Seasonal Signals**

While the lower frequency bands provide us with important insights into changes in seismic velocity at deep depths, dv/v for frequencies below 1 Hz in ocean environments is influenced by microseismic noise from ocean waves acting on the solid seafloor. Because the cross-correlation is the convolution between the noise source and the path, the microseismic noise in oceans heavily impinges upon frequencies below 1 Hz. Thus, the seasonal variability of noise sources impacts the correlation functions and, ultimately, dv/v observations. The temporal variability of noise frequency content causes apparent velocity changes due to changes in both amplitude and phase spectra resulting from waveform stretching (*62*). In stations HYSB1 and NC89, we see an annual and seasonal pattern for the dv/v time series at frequencies below 1 Hz. These observations are similar to the dv/v signals observed in frequency bands between 0.1 and 1 Hz at Axial Seamount (*63*) and the Greenland ice sheet (*64*).

While oceanographic noise source effects typically affect inter-station cross-correlations, we investigate their impact on the single-station correlation here, anticipating fewer artifacts. We utilize the open-source package Wave Model Sources of Ambient Noise (WMSAN) (*65*) to develop synthetic correlation functions for the HYSB1 station, specifically between the radial and vertical components. In particular, we selected the northern and southern Pacific Ocean basins. The notebook for the analysis is available in the CascadiaDVV project repository (src/synthetic_CC.ipynb). This workflow utilizes oceanic hindcast WAVEWATCH III data (*66*) (spectral density of the pressure field) at the ocean surface to calculate temporal variations in the source force amplitude across the Pacific Ocean's ocean bathymetry, with a spatial resolution of 0.5° x 0.5 °. By convolving these as microseism sources with the 3-D synthetic Green's function from a radially symmetric Earth, solved using AXISEM3D in ak135f model (*67*), and available at the Earthscope Consortium Syngine service. We generated synthetic seismograms for the vertical and radial components of HYSB1. Synthetic seismograms were then used to produce daily correlation functions. The correlation functions were processed using the same processing as in the data examples for the frequency band 0.1-0.3 Hz (e.g., same window length, filtering, FFT, cross-correlation, denoising, dv/v). We observe the same seasonality pattern within our synthetic DV/V time series (Fig. S3). This suggests that the noise source variability itself is likely responsible for the seasonality observed in the DV/V for frequencies less than 1 Hz.

## Hierarchical clustering of deep ETS events

We spatially selected tremor locations based on latitude bounds to focus on the region of interest, ensuring that our subsequent analysis only includes events located between 44.4°N and 44.6°N directly downdip of the RCA stations. After sorting and binning the filtered data by time to identify peaks (i.e., intervals that exceed a threshold of 200 events in a seven-day window, shown in Fig. S7), we generate 2D images of their spatial density distribution. We then extract features from these two-dimensional density maps, such as mean density, local maxima counts, and spatial moments. We normalized the feature using standard scaling to homogenize their contribution. Then, we compute a matrix with Euclidean distances between features and each event. We next apply hierarchical clustering, wherein a Ward linkage algorithm iteratively merges the two most similar clusters until all peaks coalesce into a single cluster, producing a dendrogram. Each leaf in this dendrogram represents an individual peak, and clusters merge at higher branches according to greater dissimilarity. We determine where to "cut" the dendrogram by examining silhouette scores across multiple cluster counts, selecting the threshold that yields the most coherent grouping of peaks.

We identify four clusters, defined as cluster 1 (orange), cluster 2 (green), cluster 3 (red), and cluster 4 (blue), in Figures S8 and S9. We find that cluster 3 (in green).

Our investigation also revealed that the Tremor episodes in cluster 2 also has the highest cumulative radiative energy and the highest number of recorded tremors among other events.

## Diffusivity calculation for vertical fluid migration in shallow accretionary wedge

To estimate hydraulic diffusivity for observed fluid migration along the subduction interface, we apply the classic error function solution to the diffusion equation:

$$\frac{(dv/v)}{(dv/v_0)} = erf\left(\frac{z}{2\sqrt{Dt}}\right), \quad (5)$$

where $z$ is the migration depth, $t$ is the lag time between the two depths, and $\frac{(dv/v)}{(dv/v_0)}$ is the ratio of peak dv/v values taken at the shallow over the deep values. Lag time is calculated at the lowest point in each triangular function, and as each timeseries has undergone the same processing, this will not impact the calculation of migration speeds or width comparisons between the frequencies.

From this equation, we can solve for hydraulic diffusivity $D$ as:

$$D = \frac{1}{t}\left(\frac{z}{2 \cdot erf^{-1}((dv/v)/(dv/v_0))}\right)^2 . \tag{6}$$

At Slope Base, where the fluid pulse migrated $z$ = 200 m over $t$ = 69 days, the calculated diffusivity is approximately $D = 1.4\times10^{-4}$ $m^2/s$. At Hydrate Ridge, $z$ = 300 m and $t$ = 57 days, the diffusivity is $D = 3.8\times10^{-4}$ $m^2/s$. These values are consistent with hydraulic diffusivity values in Nankai and Costa-Rica subduction zones, ranging from $10^{-6}$ and $10^{-4}$ $m^2/s$ (*68–70*).

## Fluid Flow on the Décollement or the ACF

For the **horizontal flow**, which we assume occurs along the décollement between Hydrate Ridge and Slope Base, the migration is thought to be an efficient process in a conduit. We use equation (4) of (*53*)

$$(dv/v)/(dv/v_0) = \left(\frac{1}{1+4tD/h^2}\right)^{1/2} \tag{7}$$

Given the ratio in $(dv/v)/(dv/v_0)$, assuming that channel width can range from a fault (h = 10 m) to thick underthrusted sediment (1000 m), the décollement or ACF fault width at the two stations and the lag time of 34 days, we find hydraulic diffusivity to range from 8.86 x $10^{-6}$ to 8.86 x $10^{-2}$ $m^2/s$.

We could calculate hydraulic diffusivity using the hydrogeological definition and known parameters for subduction zone shallow sediments:

$$D=k/\mu\ (\alpha+ne\ \beta) \tag{8}$$

where the intrinsic permeability is $k=1\times10^{-14}$ $m^2$, the effective porosity is $n_e$= 0.15 (*18*), the matrix compressibility is $\alpha = 1\times10^{-10}$ $Pa^{-1}$(*71*), fluid compressibility thermodynamic equation of state (TEOS-10) $\beta=4.0\times10^{-10}$ $Pa^{-1}$, and fluid viscosity $\mu=1\times10^{-3}$ Pa·s (*18*). Substituting these gives D ≈ $6.3\times10^{-2}$ $m^2/s$. This value is in the same order of magnitude as the hydraulic diffusivity calculated above for a conduit as wide as 1 km.

**Acknowledgments:**

We thank Demian Saffer and Harold Tobin for their discussion about permeability structure in a shallow subduction zone, Zoe Krauss for providing the offshore shallow tremor catalog for HYSB1, and Lisa Tomasetto for providing the WMSAN codebase used to create synthetic correlation functions.

**Funding:**

This material is based upon work supported by the Ocean Observatories Initiative (OOI), a major facility fully funded by the US National Science Foundation under Cooperative Agreement No.

2244833, and the Woods Hole Oceanographic Institution OOI Program Office. The facilities of EarthScope Consortium were used for access to waveforms, related metadata, and/or derived products used in this study. These services are funded through the National Science Foundation's Seismological Facility for the Advancement of Geoscience (SAGE) Award under Cooperative Agreement EAR-1724509 to William Wilcock. This work is supported by the Seismic Computational Platform for Empowering Discovery (SCOPED) project under the National Science Foundation (Award no. OAC-2103701 to Marine Denolle), and by the Jerome M. Paros Endowed Chair in Sensor Networks at the University of Washington (William Wilcock).

**Author contributions:**

Conceptualization: MD, MK, KFF

Methodology: MK, KFF, MD

Software: MK, KFF

Formal Analysis: MK, KFF

Validation: MK, MD, KFF

Investigation: MD, MK, KFF

Data Curation: MK

Writing – original draft: MK, MD

Writing – review & editing: MK, MD, WW, KFF

Visualization: MK

Funding acquisition: WW, MD

Project administration: MD

Supervision: WW, MD

**Competing interests:** The Authors declare that they have no competing interests.

**Data and materials availability:** All data is publicly available: the ocean bottom seismometer data from the EarthScope Data Services (https://ds.iris.edu/ds/nodes/dmc/data/) and the Ocean Observatory Initiative (https://oceanobservatories.org/cabled-array-seismometer-data/), the Wavewatch III data (*66*) is available through https://zenodo.org/records/14011562, the deep tectonic tremor catalog is provided by the Pacific Northwest Seismic Network (https://pnsn.org/tremor), the shallow tremor catalog is available (*72*), the Green's functions were available through the Syngine services (*73*). The software is also open-source, with the main project's repository available at http://doi.org/10.5281/zenodo.15334882.We used https://zenodo.org/records/11126562 for the synthetic analysis (*65*) and the open source https://doi.org/10.5281/zenodo.14534395 for the depth sensitivity.

## References

1. E. A. Wirth, V. J. Sahakian, L. M. Wallace, D. Melnick, The occurrence and hazards of great subduction zone earthquakes. *Nat. Rev. Earth Environ.* **3**, 125–140 (2022).

2. P. M. Barnes, F. C. Ghisetti, S. Ellis, J. K. Morgan, The role of protothrusts in frontal accretion and accommodation of plate convergence, Hikurangi subduction margin, New Zealand. *Geosphere* **14**, 440–468 (2018).

3. S. M. Carbotte, B. Boston, S. Han, B. Shuck, J. Beeson, J. P. Canales, H. Tobin, N. Miller, M. Nedimovic, A. Tréhu, M. Lee, M. Lucas, H. Jian, D. Jiang, L. Moser, C. Anderson, D. Judd, J. Fernandez, C. Campbell, A. Goswami, R. Gahlawat, Subducting plate structure and megathrust morphology from deep seismic imaging linked to earthquake rupture segmentation at Cascadia. *Sci. Adv.* **10**, eadl3198 (2024).

4. A. Ledeczi, M. Lucas, H. Tobin, J. Watt, N. Miller, Late Quaternary Surface Displacements on Accretionary Wedge Splay Faults in the Cascadia Subduction Zone: Implications for Megathrust Rupture. *Seismica* **2** (2023).

5. T. Lay, Why giant earthquakes keep catching us out. *Nature* **483**, 149–150 (2012).

6. E. O. Lindsey, R. Mallick, J. A. Hubbard, K. E. Bradley, R. V. Almeida, J. D. P. Moore, R. Bürgmann, E. M. Hill, Slip rate deficit and earthquake potential on shallow megathrusts. *Nat. Geosci.* **14**, 321–326 (2021).

7. W. M. Behr, R. Bürgmann, What's down there? The structures, materials and environment of deep-seated slow slip and tremor. *Philos. Trans. R. Soc. A*, doi: 10.1098/rsta.2020.0218 (2021).

8. R. Bürgmann, The geophysics, geology and mechanics of slow fault slip. *Earth Planet. Sci. Lett.* **495**, 112–134 (2018).

9. M. Nakano, T. Hori, E. Araki, S. Kodaira, S. Ide, Shallow very-low-frequency earthquakes accompany slow slip events in the Nankai subduction zone. *Nat. Commun.* **9**, 984 (2018).

10. R. Plata-Martinez, S. Ide, M. Shinohara, E. S. Garcia, N. Mizuno, L. A. Dominguez, T. Taira, Y. Yamashita, A. Toh, T. Yamada, J. Real, A. Husker, V. M. Cruz-Atienza, Y. Ito, Shallow slow earthquakes to decipher future catastrophic earthquakes in the Guerrero seismic gap. *Nat. Commun.* **12**, 3976 (2021).

11. K. Woods, L. M. Wallace, C. A. Williams, I. J. Hamling, S. C. Webb, Y. Ito, N. Palmer, R. Hino, S. Suzuki, M. K. Savage, E. Warren-Smith, K. Mochizuki, Spatiotemporal Evolution of Slow Slip Events at the Offshore Hikurangi Subduction Zone in 2019 Using GNSS, InSAR, and Seafloor Geodetic Data. *J. Geophys. Res. Solid Earth* **129**, e2024JB029068 (2024).

12. M. A. L. Walton, L. M. Staisch, T. Dura, J. K. Pearl, B. Sherrod, J. Gomberg, S. Engelhart, A. Tréhu, J. Watt, J. Perkins, R. C. Witter, N. Bartlow, C. Goldfinger, H. Kelsey, A. E. Morey, V. J. Sahakian, H. Tobin, K. Wang, R. Wells, E. Wirth, Toward an Integrative Geological and Geophysical View of Cascadia Subduction Zone Earthquakes. *Annu. Rev. Earth Planet. Sci.* **49**, 367–398 (2021).

13. C. Goldfinger, C. H. Nelson, A. E. Morey, J. E. Johnson, J. R. Patton, E. B. Karabanov, J. Gutierrez-Pastor, A. T. Eriksson, E. Gracia, G. Dunhill, R. J. Enkin, A. Dallimore, T. Vallier, "Turbidite event history—Methods and implications for Holocene paleoseismicity of the Cascadia subduction zone" (1661-F, U.S. Geological Survey, 2012); https://doi.org/10.3133/pp1661F.

14. G. M. Schmalzle, R. McCaffrey, K. C. Creager, Central Cascadia subduction zone creep. *Geochem. Geophys. Geosystems* **15**, 1515–1532 (2014).

15. C. Goldfinger, L. D. Kulm, R. S. Yeats, L. McNeill, C. Hummon, Oblique strike-slip faulting of the central Cascadia submarine forearc. *J. Geophys. Res. Solid Earth* **102**, 8217–8243 (1997).

16. S. Han, N. L. Bangs, S. M. Carbotte, D. M. Saffer, J. C. Gibson, Links between sediment consolidation and Cascadia megathrust slip behaviour. *Nat. Geosci.* **10**, 954–959 (2017).

17. R. D. Hyndman, P. A. McCrory, A. Wech, H. Kao, J. Ague, Cascadia subducting plate fluids channelled to fore-arc mantle corner: ETS and silica deposition. *J. Geophys. Res. Solid Earth* **120**, 4344–4358 (2015).

18. B. T. Philip, E. A. Solomon, D. S. Kelley, A. M. Tréhu, T. L. Whorley, E. Roland, M. Tominaga, R. W. Collier, Fluid sources and overpressures within the central Cascadia Subduction Zone revealed by a warm, high-flux seafloor seep. *Sci. Adv.* **9**, eadd6688 (2023).

19. B. M. A. Teichert, M. E. Torres, G. Bohrmann, A. Eisenhauer, Fluid sources, fluid pathways and diagenetic reactions across an accretionary prism revealed by Sr and B geochemistry. *Earth Planet. Sci. Lett.* **239**, 106–121 (2005).

20. J. M. Gosselin, P. Audet, C. Estève, M. McLellan, S. G. Mosher, A. J. Schaeffer, Seismic evidence for megathrust fault-valve behavior during episodic tremor and slip. *Sci. Adv.* **6**, eaay5174 (2020).

21. T. Tonegawa, S. Takemura, S. Yabe, K. Yomogida, Fluid Migration Before and During Slow Earthquakes in the Shallow Nankai Subduction Zone. *J. Geophys. Res. Solid Earth* **127**, e2021JB023583 (2022).

22. K. Okubo, B. G. Delbridge, M. A. Denolle, Monitoring Velocity Change Over 20 Years at Parkfield. *J. Geophys. Res. Solid Earth* **129**, e2023JB028084 (2024).

23. T. Clements, M. A. Denolle, The Seismic Signature of California's Earthquakes, Droughts, and Floods. *J. Geophys. Res. Solid Earth* **128**, e2022JB025553 (2023).

24. G. Hillers, L. Retailleau, M. Campillo, A. Inbal, J.-P. Ampuero, T. Nishimura, In situ observations of velocity changes in response to tidal deformation from analysis of the high-frequency ambient wavefield. *J. Geophys. Res. Solid Earth* **120**, 210–225 (2015).

25. T. Takano, T. Nishimura, H. Nakahara, H. Ueda, E. Fujita, Sensitivity of Seismic Velocity Changes to the Tidal Strain at Different Lapse Times: Data Analyses of a Small Seismic Array at Izu-Oshima Volcano. *J. Geophys. Res. Solid Earth* **124**, 3011–3023 (2019).

26. C. Sens-Schönfelder, T. Eulenfeld, Probing the in situ Elastic Nonlinearity of Rocks with Earth Tides and Seismic Noise. *Phys. Rev. Lett.* **122**, 138501 (2019).

27. E. Larose, S. Carrière, C. Voisin, P. Bottelin, L. Baillet, P. Guéguen, F. Walter, D. Jongmans, B. Guillier, S. Garambois, F. Gimbert, C. Massey, Environmental seismology: What can we learn on earth surface processes with ambient noise? *J. Appl. Geophys.* **116**, 62–74 (2015).

28. S. Li, K. Wang, Y. Wang, Y. Jiang, S. E. Dosso, Geodetically Inferred Locking State of the Cascadia Megathrust Based on a Viscoelastic Earth Model. *J. Geophys. Res. Solid Earth* **123**, 8056–8072 (2018).

29. E. E. Davis, T. Sun, K. Becker, M. Heesemann, H. Villinger, K. Wang, Deep-sea borehole fluid pressure and temperature observations at subduction zones and their geodynamic implications. *Can. J. Earth Sci.*, doi: 10.1139/cjes-2024-0093 (2024).

30. R. Snieder, C. Sens-Schönfelder, R. Wu, The time dependence of rock healing as a universal relaxation process, a tutorial. *Geophys. J. Int.* **208**, 1–9 (2017).

31. E. Araki, D. M. Saffer, A. J. Kopf, L. M. Wallace, T. Kimura, Y. Machida, S. Ide, E. Davis, IODP Expedition 365 shipboard scientists, Recurring and triggered slow-slip events near the trench at the Nankai Trough subduction megathrust. *Science* **356**, 1157–1160 (2017).

32. Z. Krauss, W. S. D. Wilcock, K. C. Creager, Potential Shallow Tectonic Tremor Signals Near the Deformation Front in Central Cascadia. doi: 10.22541/au.173498202.28863970/v1 (2024).

33. A. G. Wech, Interactive Tremor Monitoring. *Seismol. Res. Lett.* **81**, 664–669 (2010).

34. G. Hillers, S. Husen, A. Obermann, T. Planès, E. Larose, M. Campillo, Noise-based monitoring and imaging of aseismic transient deformation induced by the 2006 Basel reservoir stimulation. *Geophysics* **80**, KS51–KS68 (2015).

35. V. M. Cruz-Atienza, C. Villafuerte, H. S. Bhat, Rapid tremor migration and pore-pressure waves in subduction zones. *Nat. Commun.* **9**, 2900 (2018).

36. A. Ghosh, J. E. Vidale, J. R. Sweet, K. C. Creager, A. G. Wech, H. Houston, E. E. Brodsky, Rapid, continuous streaking of tremor in Cascadia. *Geochem. Geophys. Geosystems* **11** (2010).

37. I. Stone, J. E. Vidale, S. Han, E. Roland, Catalog of Offshore Seismicity in Cascadia: Insights Into the Regional Distribution of Microseismicity and its Relation to Subduction Processes. *J. Geophys. Res. Solid Earth* **123**, 641–652 (2018).

38. R. J. Burgette, R. J. Weldon II, D. A. Schmidt, Interseismic uplift rates for western Oregon and along-strike variation in locking on the Cascadia subduction zone. *J. Geophys. Res. Solid Earth* **114** (2009).

39. K. Chaudhuri, A. Ghosh, Widespread Very Low Frequency Earthquakes (VLFEs) Activity Offshore Cascadia. *Geophys. Res. Lett.* **49**, e2022GL097962 (2022).

40. J. Nakajima, N. Uchida, Repeated drainage from megathrusts during episodic slow slip. *Nat. Geosci.* **11**, 351–356 (2018).

41. J. L. Rubinstein, M. La Rocca, J. E. Vidale, K. C. Creager, A. G. Wech, Tidal Modulation of Nonvolcanic Tremor. *Science* **319**, 186–189 (2008).

42. J. L. Rubinstein, J. Gomberg, J. E. Vidale, A. G. Wech, H. Kao, K. C. Creager, G. Rogers, Seismic wave triggering of nonvolcanic tremor, episodic tremor and slip, and earthquakes on Vancouver Island. *J. Geophys. Res. Solid Earth* **114** (2009).

43. A. J. Calvert, M. G. Bostock, G. Savard, M. J. Unsworth, Cascadia low frequency earthquakes at the base of an overpressured subduction shear zone. *Nat. Commun.* **11**, 3874 (2020).

44. P. Audet, M. G. Bostock, N. I. Christensen, S. M. Peacock, Seismic evidence for overpressured subducted oceanic crust and megathrust fault sealing. *Nature* **457**, 76–78 (2009).

45. S. Ozawa, Y. Yang, E. M. Dunham, Fault-Valve Instability: A Mechanism for Slow Slip Events. *J. Geophys. Res. Solid Earth* **129**, e2024JB029165 (2024).

46. P. Herath, P. Audet, Fluid upwelling across the Hikurangi subduction thrust during deep slow-slip earthquakes. *Commun. Earth Environ.* **5**, 1–8 (2024).

47. R. C. Viesca, D. I. Garagash, Ubiquitous weakening of faults due to thermal pressurization. *Nat. Geosci.* **8**, 875–879 (2015).

48. E. Warren-Smith, B. Fry, L. Wallace, E. Chon, S. Henrys, A. Sheehan, K. Mochizuki, S. Schwartz, S. Webb, S. Lebedev, Episodic stress and fluid pressure cycling in subducting oceanic crust during slow slip. *Nat. Geosci.* **12**, 475–481 (2019).

49. S. Kita, H. Houston, S. Yabe, S. Tanaka, Y. Asano, T. Shibutani, N. Suda, Effects of episodic slow slip on seismicity and stress near a subduction-zone megathrust. *Nat. Commun.* **12**, 7253 (2021).

50. G. D. Egbert, B. Yang, P. A. Bedrosian, K. Key, D. W. Livelybrooks, A. Schultz, A. Kelbert, B. Parris, Fluid transport and storage in the Cascadia forearc influenced by overriding plate lithology. *Nat. Geosci.* **15**, 677–682 (2022).

51. H. Tobin, J. C. Moore, M. E. MacKay, D. L. Orange, L. Kulm, Fluid flow along a strike-slip fault at the toe of the Oregon accretionary prism: Implications for the geometry of frontal accretion. *GSA Bull.* **105**, 569–582 (1993).

52. J. R. Delph, A. Levander, F. Niu, Fluid Controls on the Heterogeneous Seismic Characteristics of the Cascadia Margin. *Geophys. Res. Lett.* **45**, 11,021-11,029 (2018).

53. D. M. Saffer, The permeability of active subduction plate boundary faults. *Geofluids* **15**, 193–215 (2014).

54. E. E. Davis, T. Sun, M. Heesemann, K. Becker, A. Schlesinger, Long-Term Offshore Borehole Fluid-Pressure Monitoring at the Northern Cascadia Subduction Zone and Inferences Regarding the State of Megathrust Locking. *Geochem. Geophys. Geosystems* **24**, e2023GC010910 (2023).

55. J. C. Moore, G. F. Moore, G. R. Cochrane, H. J. Tobin, Negative-polarity seismic reflections along faults of the Oregon accretionary prism: Indicators of overpressuring. *J. Geophys. Res. Solid Earth* **100**, 12895–12906 (1995).

56. G. R. Cochrane, J. C. Moore, M. E. MacKay, G. F. Moore, Velocity and inferred porosity model of the Oregon accretionary prism from multichannel seismic reflection data: Implications on sediment dewatering and overpressure. *J. Geophys. Res. Solid Earth* **99**, 7033–7043 (1994).

57. Materials and methods are available as supplementary materials.

58. T. Clements, M. A. Denolle, SeisNoise.jl: Ambient Seismic Noise Cross Correlation on the CPU and GPU in Julia. *Seismol. Res. Lett.* **92**, 517–527 (2020).

59. L. Moreau, L. Stehly, P. Boué, Y. Lu, E. Larose, M. Campillo, Improving ambient noise correlation functions with an SVD-based Wiener filter. *Geophys. J. Int.* **211**, 418–426 (2017).

60. C. Sens-Schönfelder, E. Pomponi, A. Peltier, Dynamics of Piton de la Fournaise volcano observed by passive image interferometry with multiple references. *J. Volcanol. Geotherm. Res.* **276**, 32–45 (2014).

61. K. Luu, disba: Numba-accelerated computation of surface wave dispersion, version v0.7.0, Zenodo (2024); https://doi.org/10.5281/zenodo.14534395.

62. Z. Zhan, V. C. Tsai, R. W. Clayton, Spurious velocity changes caused by temporal variations in ambient noise frequency content. *Geophys. J. Int.* **194**, 1574–1581 (2013).

63. M. K. Lee, "Imaging and characterizing subseafloor structures associated with active magmatic and hydrothermal processes at and near seamounts on the Juan de Fuca plate from ridge to trench," thesis, Columbia University (2024).

64. B. Luo, S. Zhang, H. Zhu, Monitoring Seasonal Fluctuation and Long-Term Trends for the Greenland Ice Sheet Using Seismic Noise Auto-Correlations. *Geophys. Res. Lett.* **50**, e2022GL102146 (2023).

65. L. Tomasetto, P. Boué, F. Ardhuin, E. Stutzman, Z. Xu, R. de Plaen, L. Stehly, F. Ardhuin, E. Stutzman, Z. Xu, R. de Plaen, L. Stehly, WMSAN Python Package: From Oceanic Forcing to Synthetic Cross-correlations of Microseismic Noise. *EathArXiv*, doi: 10.31223/X5CB08 (2024).

66. A. Gaffet, X. Bertin, D. Sous, H. Michaud, A. Roland, A new global high resolution model for the tropical ocean using WAVEWATCH III version 7.14 - WaveWatchIII codebase. (2024).

67. L. Krischer, A. R. Hutko, M. van Driel, S. Stähler, M. Bahavar, C. Trabant, T. Nissen-Meyer, On-Demand Custom Broadband Synthetic Seismograms. *Seismol. Res. Lett.* **88**, 1127–1140 (2017).

68. A. L. LaBonte, K. M. Brown, Y. Fialko, Hydrologic detection and finite element modeling of a slow slip event in the Costa Rica prism toe. *J. Geophys. Res. Solid Earth* **114** (2009).

69. S. B. Hammerschmidt, E. E. Davis, A. Hüpers, A. Kopf, Limitation of fluid flow at the Nankai Trough megasplay fault zone. *Geo-Mar. Lett.* **33**, 405–418 (2013).

70. W. Tanikawa, H. Mukoyoshi, W. Lin, T. Hirose, A. Tsutsumi, Pressure dependence of fluid transport properties of shallow fault systems in the Nankai subduction zone. *Earth Planets Space* **66**, 90 (2014).

71. B. Dugan, T. C. Sheahan, Offshore sediment overpressures of passive margins: Mechanisms, measurement, and models. *Rev. Geophys.* **50** (2012).

72. Z. Krauss, zoekrauss/obs_tremor: Submission, version Submission, Zenodo (2024); https://doi.org/10.5281/zenodo.14532861.

73. T. Nissen-Meyer, M. van Driel, S. C. Stähler, K. Hosseini, S. Hempel, L. Auer, A. Colombi, A. Fournier, AxiSEM: broadband 3-D seismic wavefields in axisymmetric media. *Solid Earth* **5**, 425–445 (2014).

# Supplementary Materials for

## Active Protothrusts and Fluid Highways: Seismic Noise Reveals Hidden Subduction Dynamics in Cascadia

**Authors:** Maleen Kidiwela, Marine A. Denolle, William S. D. Wilcock, Kuan-Fu Feng

Corresponding author: Maleen Kidiwela, seismic@uw.edu

**The PDF file includes:**

Figs. S1 to S10
Tables S1 to S2

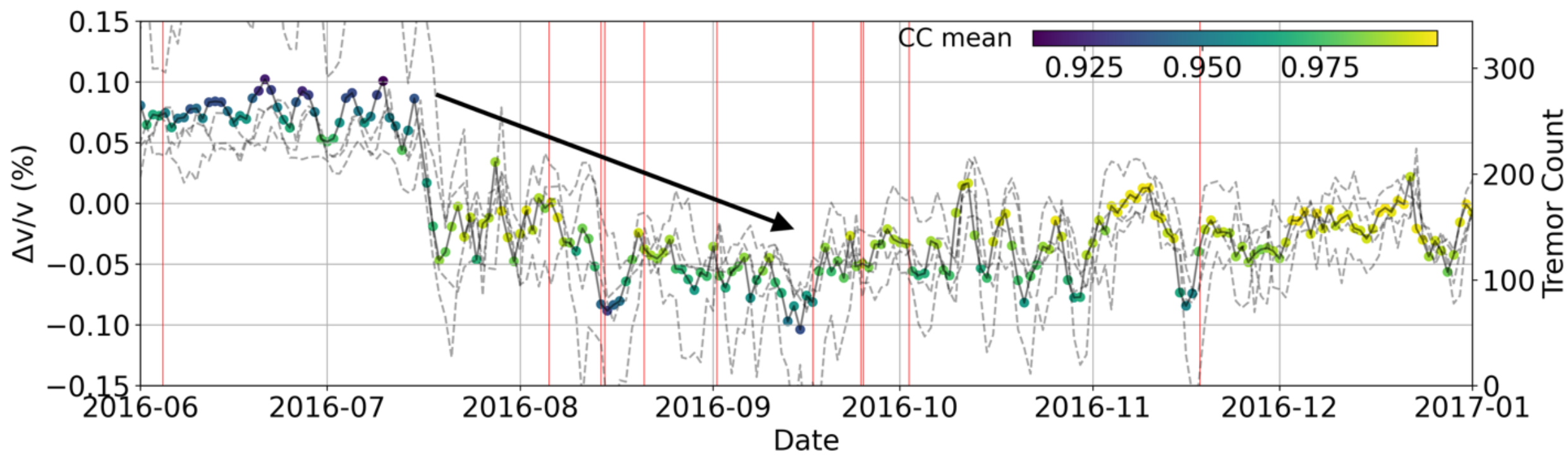

**Figure S1: dv/v of HYSB1 at 3-5 Hz under a 3-day stacked 2016 slow slip event, weighted averaging ZE and ZN positive/negative components.** The arrow indicates the slowly declining dv/v over two months to its lowest dv/v value.

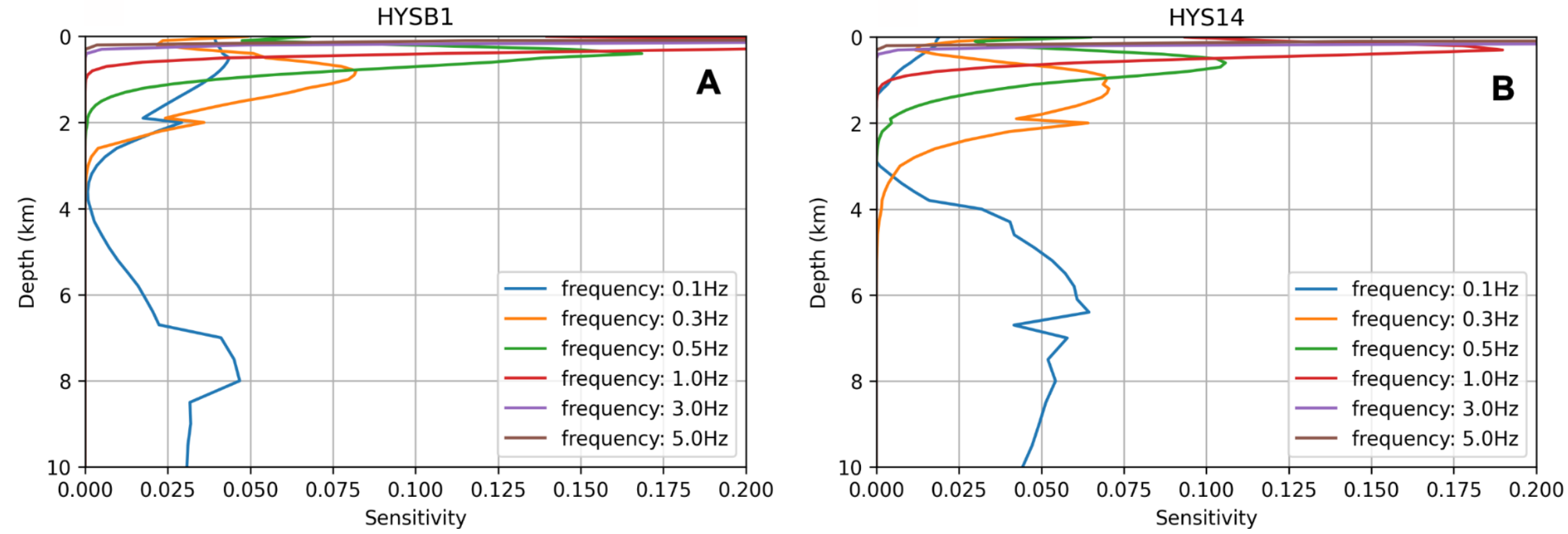

**Figure S2: Depth sensitivity of the dv/v measurements** (assuming Rayleigh waves dominated) as a function of frequency and depth for HYSB1 in (**A**) and HYS14 in (**B**).

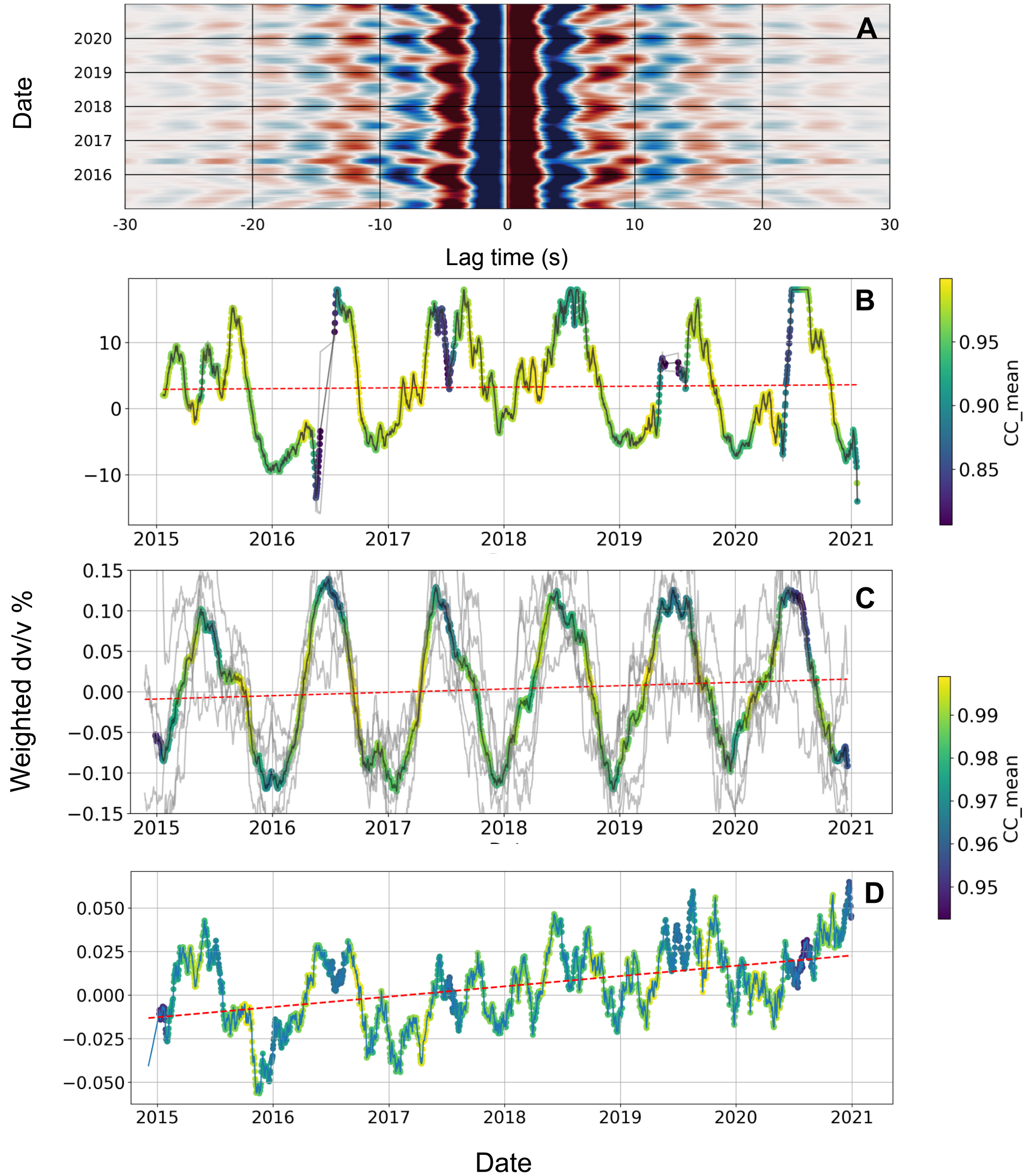

**Figure S3: Low frequency observed and synthetic cross-correlation and dv/v.** (**A**) Synthetic cross-correlation modeled at HYSB1, filtered 0.1-0.3 Hz, using microseismic signals from wavewatch data and WMSAN to compute. (**B**) synthetic dv/v obtained from (A). (**C**) Observed dv/v at HYSB in the 0.1-0.3 Hz range, largely dominated by seasonal spurious signals. (**D**) Notched-filter time series of (C) to remove the 200-400 day period and explore the residual and the remaining positive trend illustrating a compression.

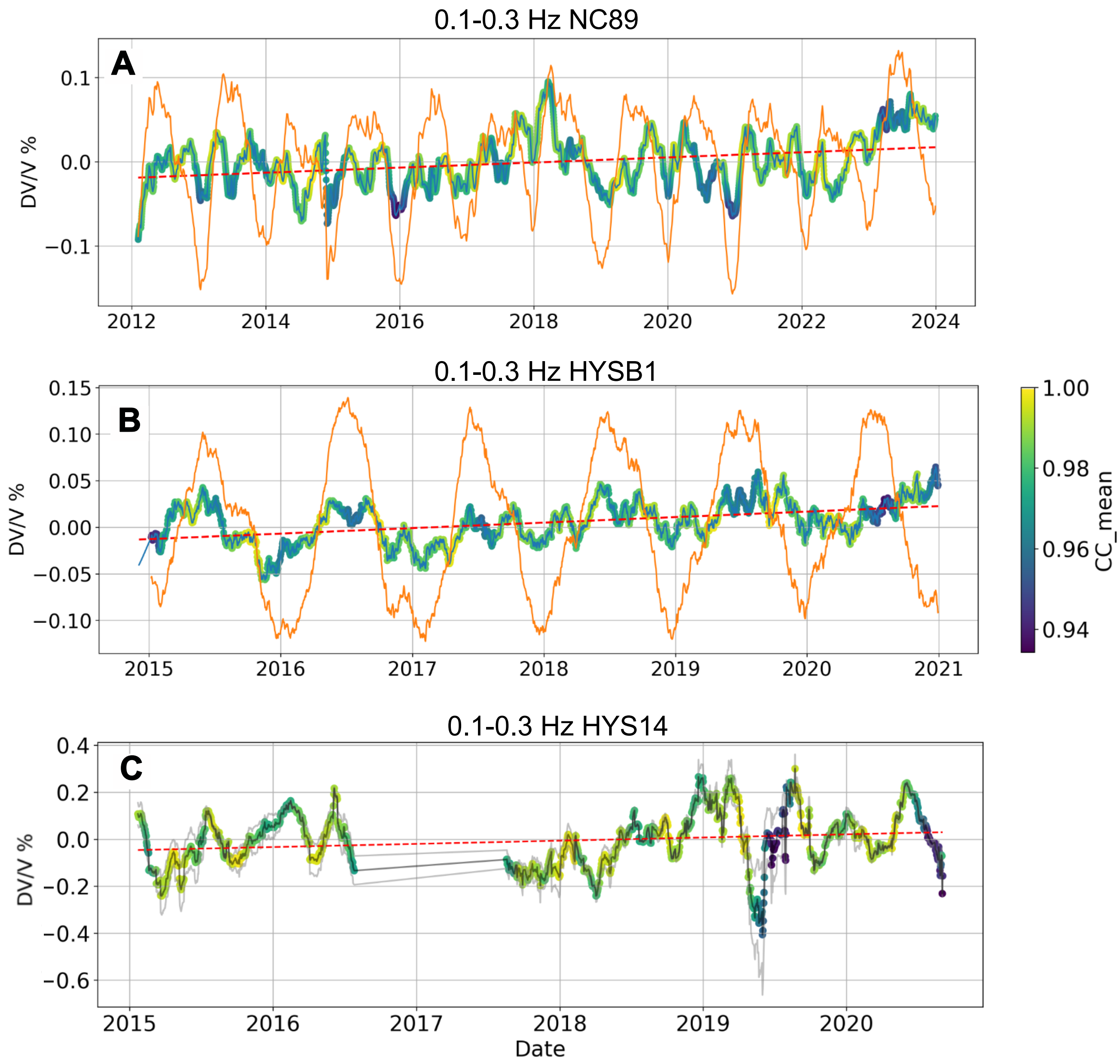

**Figure S4: Removing seasonality from low-frequency dv/v.** We removed seasonal signals using a Gaussian notch filter in the 0.1-0.3 Hz dv/v in (**A**) NC89, (**B**) HYSB1, and (**C**) HYS14. Color-coded values have the cumulative correlation coefficient after stretching.

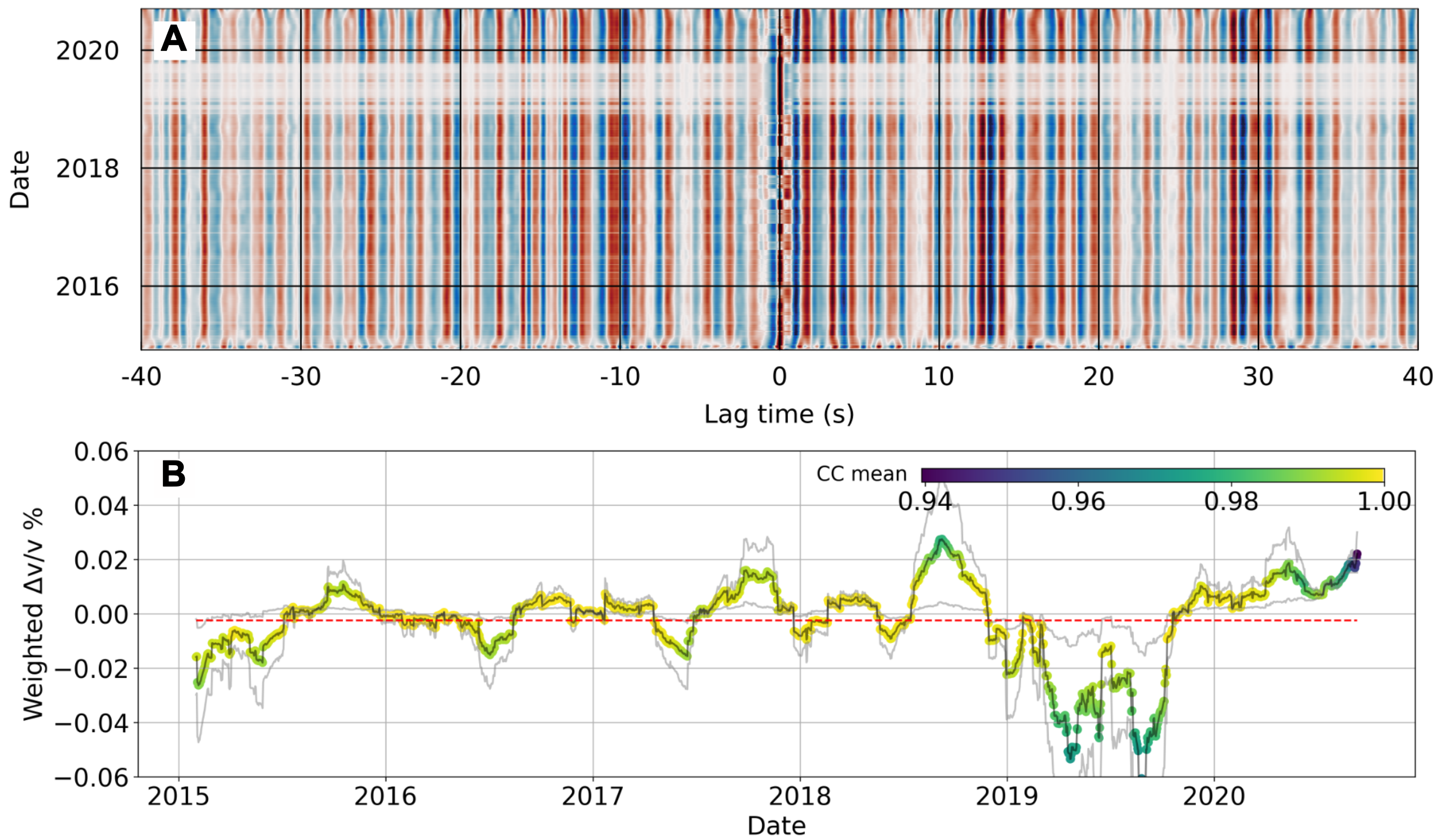

**Figure S5**: **Interstation dv/v between HYS14 and HYSB1** in the 0.5-2 Hz frequency band, with (**A**) the vertical-to-vertical cross-correlation, and (**B**) the resulting dv/v, color-coded by the cumulative correlation coefficient. The dv/v values are 10 times lower than the single-station measurements, which is expected from the decreased sensitivity between the stations. Nevertheless, this dv/v time series coincides well with the average of the two single-station dv/v time series, demonstrating that the transient is located at and between the two stations.

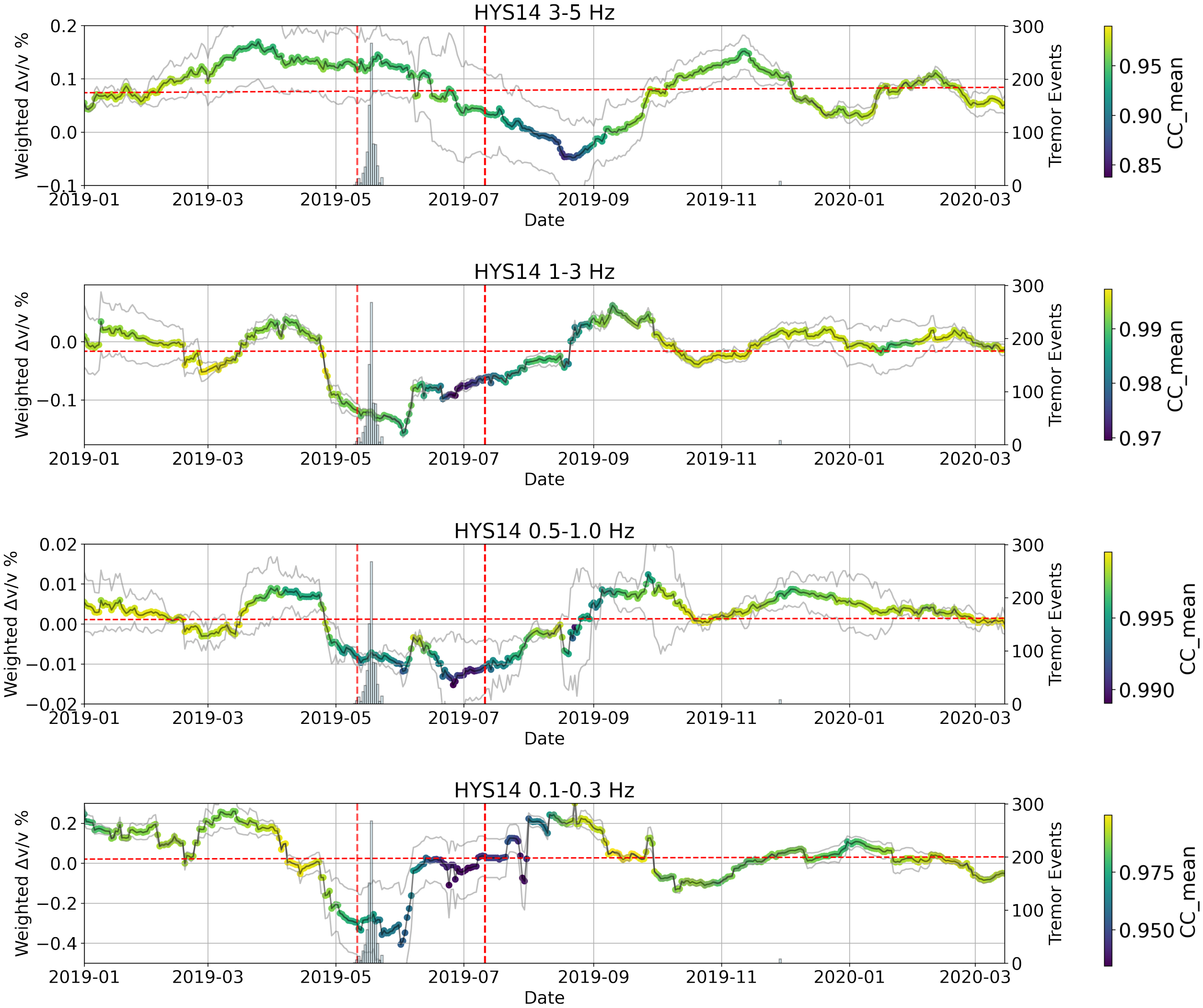

**Figure S6**: **Multi-frequency weighted dv/v at HYS14 for the 2019 transient event**. From high frequency (3-5 Hz, top panel) to low frequency (0.1-0.3 Hz, bottom panel), sorted by depth sensitivity. The cross-correlations have been averaged over 40 days. dv/v is color-coded by the cumulative correlation coefficient after stretching. We observe a shift from deep dv/v to be early, and shallow dv/v to be later. Note that Krauss' shallow tremors (*27*) also appear during the transient.

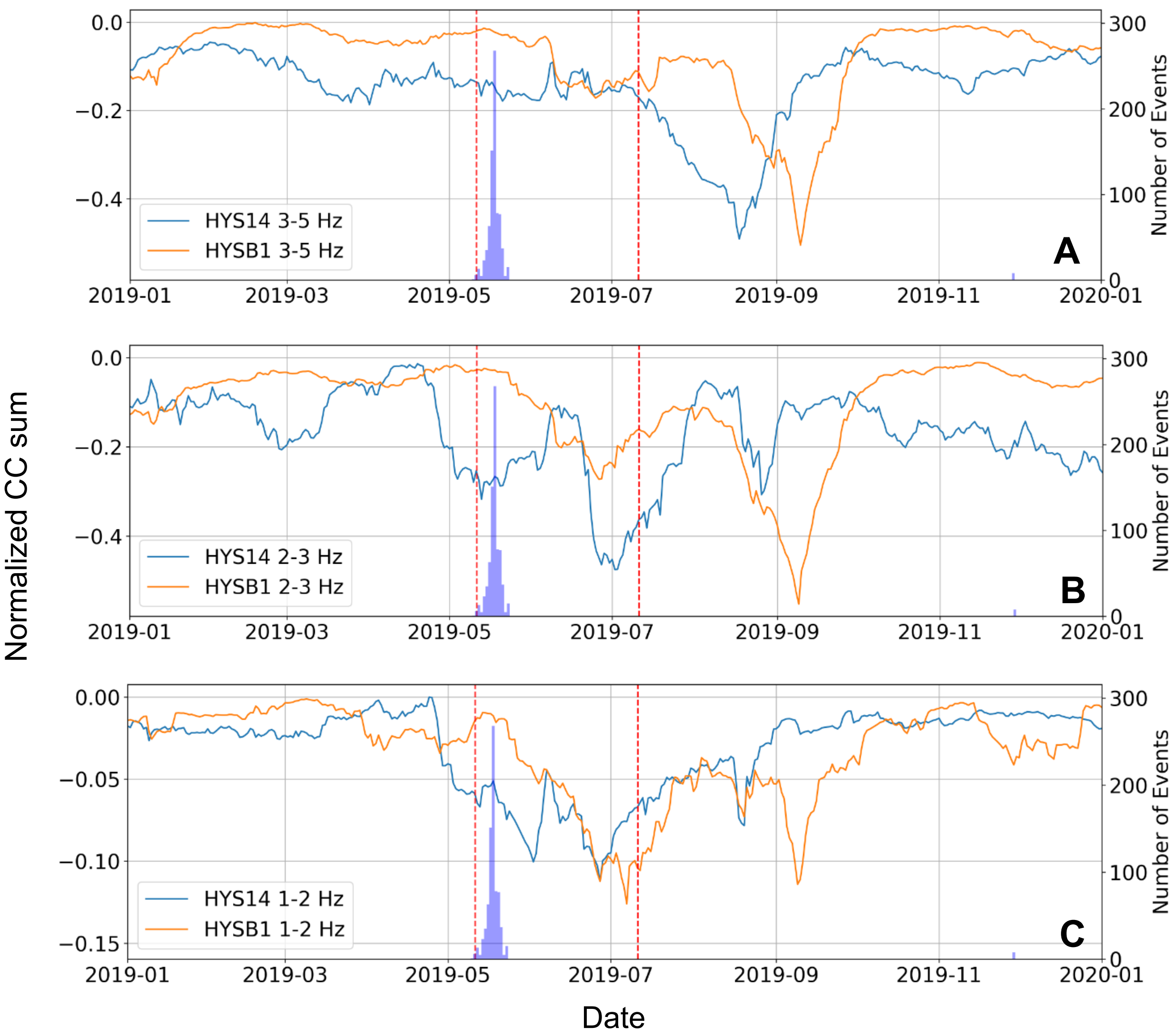

**Figure S7**: **Normalized sum of squared correlation coefficients of each time series** for (**A**) 3-5 Hz (**B**) 2-3 Hz (**C**) 1-2 Hz in HYS14 and HYSB1 stations during the 2019 ETS event. The squared sum of correlation coefficients of each time series are normalized to their maximum value and shifted to align the maximum values with zero for improved visualization.

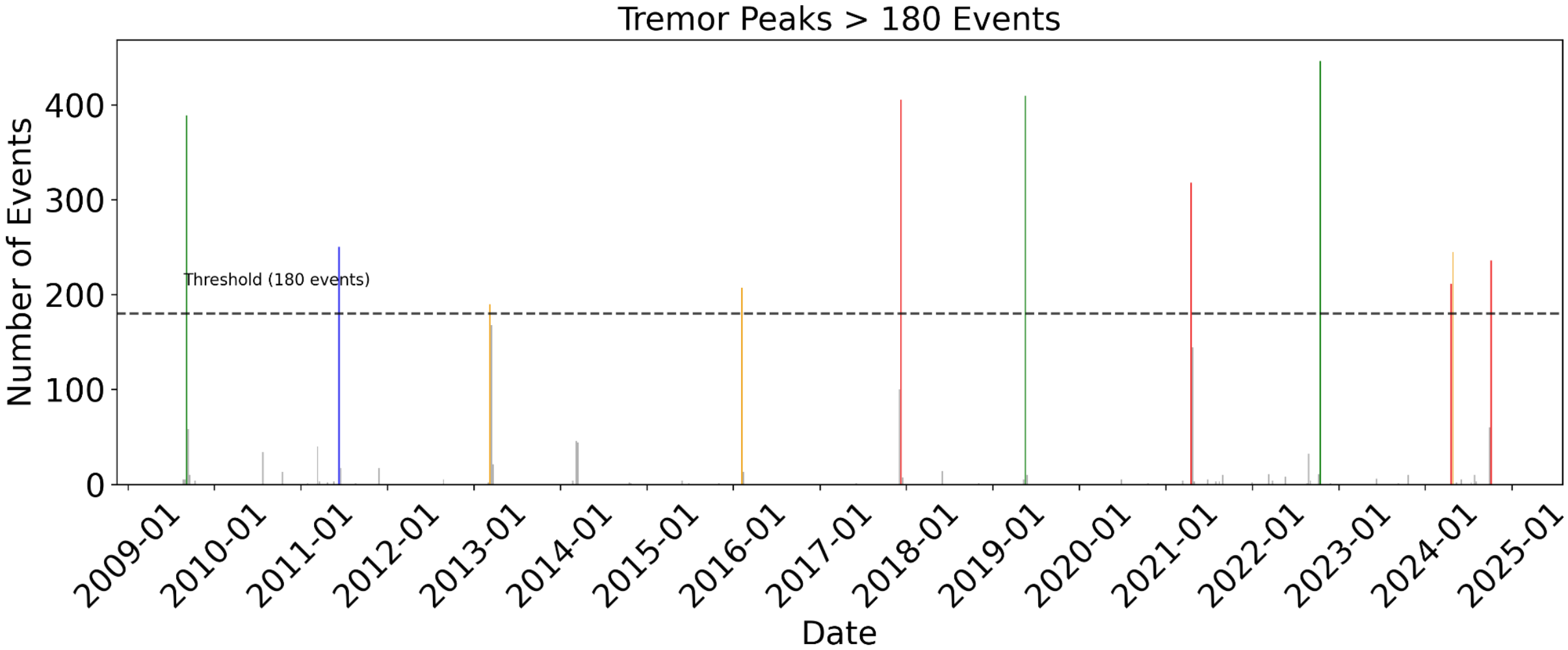

**Figure S8: 7-day binned tremor counts for 44.4°N-44.6°N latitude range.** Colored events refer to those exceeding 180 event counts per episode and correspond to the dendrogram shown in Figure S9.

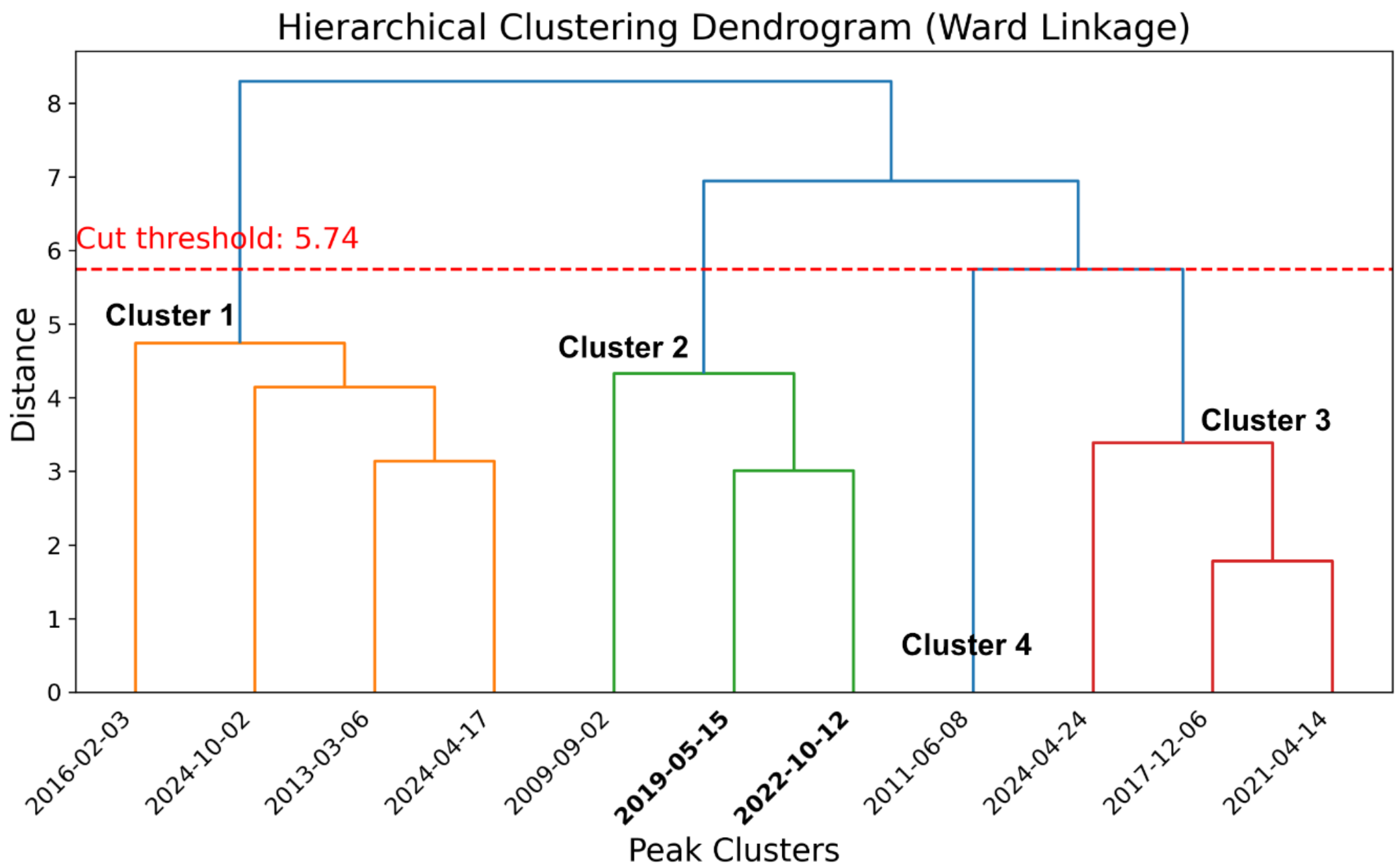

**Figure S9: Ward linkage dendrogram of clustered tremor events.** Bolded events are the two events with dv/v drops in HYSB1.

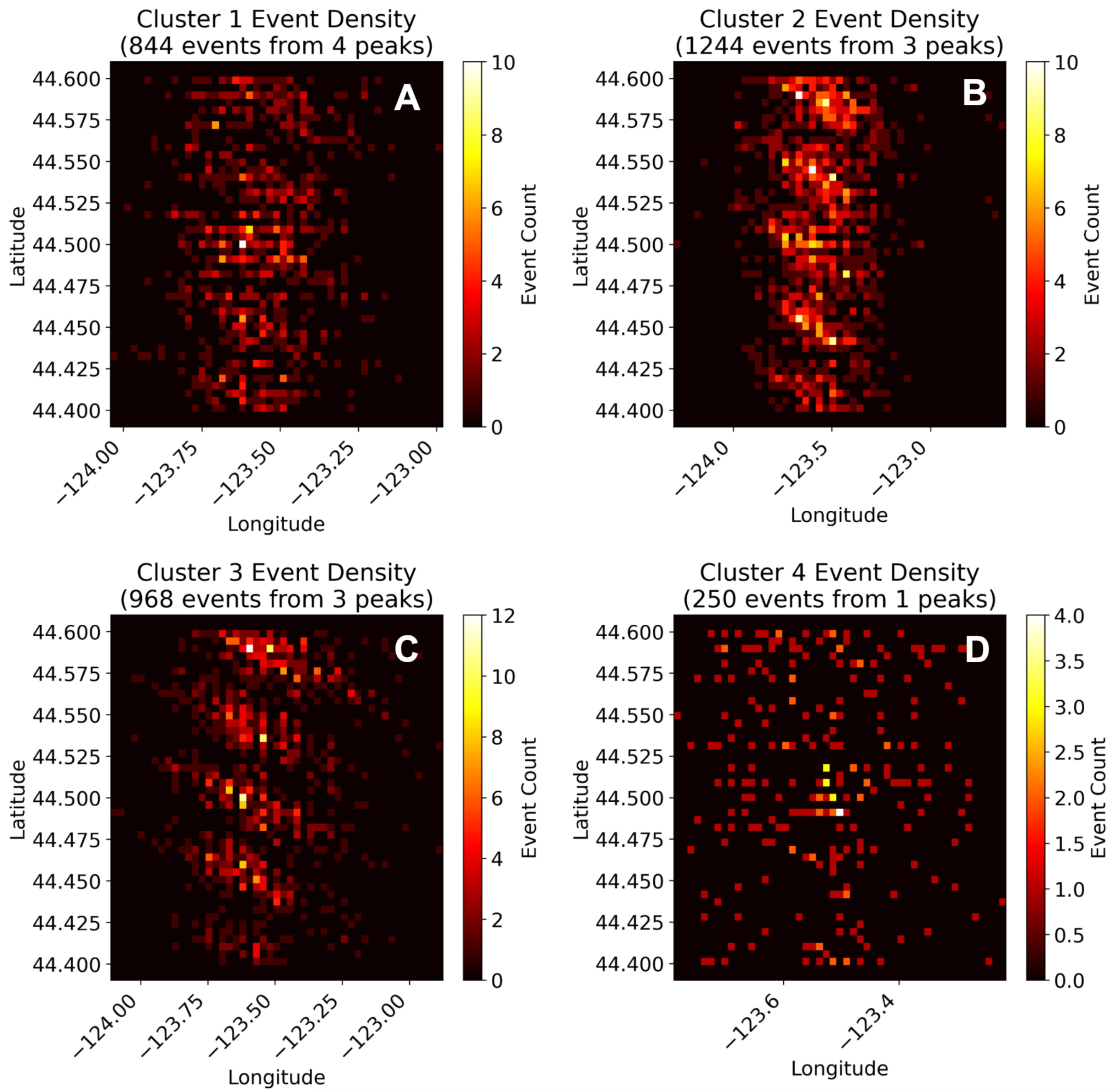

**Figure S10: Density map for each cluster of events in the dendrogram of ward linkage by summing all events within the same cluster.** The cluster density plots correspond to the clusters shown in Figure S9, specifically (**A**) the orange cluster, (**B**) the green cluster, (**C**) the red cluster, and (**D**) the blue cluster in the dendrogram. Cluster B is associated with the dv/v drops observed.

| Freq band (Hz) | 0.1-0.3 | 0.3-0.6 | 0.5-1 | 1-3 | 3-5 |
|---|---|---|---|---|---|
| Coda window: start and end times in seconds. | 10-60 | 10-30 | 10-20 | 2-10 | 2-10 |

**Table S1: Choice of frequency band and coda window.**

| **Frequency** | $\boldsymbol{H_d}$ **at HYSB1 (km)** | $\boldsymbol{H_d}$ **at HYS14 (km)** |
|---|---|---|
| **5 Hz** | 0.03 km | 0.04 km |
| **3 Hz** | 0.1 km | 0.1 km |
| **1 Hz** | 0.2 km | 0.3 km |
| **0.5 Hz** | 0.5 km | 0.75 km |
| **0.3 Hz** | 1 km | 1.5 km |
| **0.1 Hz** | 8 km | 8 km |

**Table S2: Frequency-dependent depth sensitivity of HYSB1 and HYS14**.